\documentclass[aps,twocolumn,pra,amsmath,amssymb,aps]{revtex4-1}
\usepackage{graphicx}
\usepackage{dcolumn}
\usepackage{bm}
\usepackage{graphicx,lastpage}
\usepackage{graphics}
\usepackage{amsmath}
\usepackage{amssymb}
\usepackage{upgreek}
\usepackage{color}
\usepackage{float}
\usepackage{soul}
\usepackage{ragged2e}
\usepackage{subfig,caption,subcaption}
\usepackage{censor}
\usepackage{enumerate}
\usepackage{textcomp}
\usepackage{braket}
\usepackage{placeins}
\usepackage{ulem}
\usepackage{xr}
\begin{document}

\title{Ergotropy of a Photosynthetic Reaction Center}

\author{Trishna Kalita$^\dag$}
\author{Manash Jyoti Sarmah$^\dag$}
\author{Javed Akhtar}
%\author{Bitap Raj Thakuria}
\author{Himangshu Prabal Goswami}
\email{hpg@gauhati.ac.in, $^\dag$Equal Contributors}

\affiliation{QuAInT Research Group, Department of Chemistry, Gauhati University, Guwahati–781014, Assam, India}

\date{\today}

\begin{abstract}
We theoretically analyze the Photosystem II reaction center using a quantum master equation approach, where excitonic and charge-transfer rates are computed at the Redfield and Förster levels with realistic spectral densities. The focus is on \textit{ergotropy}, the maximum work extractable from a quantum state without energy loss. We compute the ergotropy by constructing passive states in the thermodynamic sense. Among the electron transfer pathways, those involving charge separation between $Chl_{D1}$ and $Phe_{D1}$, as well as a route passing through three sequential charge-separated states, yield higher ergotropy, suggesting greater capacity for work extraction, akin to quantum energy capacitors. A third pathway, bypassing the $Chl_{D1},Phe_{D1}$ pair, shows significantly reduced ergotropy. These differences arise from population-induced transitions between active and passive regimes. Our findings highlight how biological systems may exploit non-equilibrium population structures to optimize energy conversion, connecting quantum thermodynamic principles to biological energy harvesting.

\end{abstract}

\maketitle

\section{\label{sec:level1}Introduction}
Ergotropy quantifies the maximal work extractable from a quantum state through unitary evolution, distinguishing between energy that can be converted into work and energy that, while stored, remains inaccessible under unitary dynamics \cite{Allahverdyan2004}. This distinction is relevant for nonequilibrium quantum states, where both active and passive states coexist \cite{francica2020quantum}. Passive states, including thermal states, are defined by their inability to yield work through any unitary operation indicating deviations from thermal equilibrium \cite{perarnau2015most}. Conversely, active states exhibit nonzero ergotropy and enable efficient energy-to-work conversion. Ergotropy thus refines classical notions of work by characterizing energy conversion in quantum devices such as batteries, engines, and diodes \cite{vinjanampathy2016quantum, alicki2018introduction, binder2015quantum, andolina2019quantum}, and has been recorded in feedback experiments \cite{joshi2024maximal} and spin systems\cite{niu2024experimental}.

Thermodynamically, the passive or active nature of a state is determined by population ordering in the system Hamiltonian’s eigenbasis \cite{francica2020quantum}. Passive states show monotonically decreasing populations with energy. Unitary transformations that preserve eigenvalues reveal the ergotropic potential \cite{touil2021ergotropy,PhysRevA.110.032213} are reversible and conserve the spectrum. These form the theoretical foundation of ergotropy in systems out of equilibrium \cite{allahverdyan2004maximal}. At the nanoscale, nonequilibrium conditions and quantum features such as coherence and entanglement can be exploited to extract maximal work, suggesting that biological systems may utilize quantum resources not just for energy storage but for efficient work extraction. This links quantum thermodynamics with bioenergetics and opens possibilities for novel modes of energy conversion in biological settings. A notable example is the Photosystem II Reaction Center (PSIIRC), which embodies this quantum thermodynamic–biological interplay \cite{Ferreira2004}.
It is a protein-pigment complex central to photosynthetic charge separation facilitating energy transfer via multiple quantum transport pathways \cite{blankenship2014molecular, scholes2011lessons, van2000photosynthetic, engel2007evidence,STONES20176871,Novoderezhkin2011}. Upon photoexcitation, excitons propagate along these pathways, inducing charge separation through coherent quantum dynamics \cite{ishizaki2009theoretical, hoki2009quantum, fassioli2010quantum}. The protein environment plays a dual role by modulating excitonic couplings and imposing spatial and energetic asymmetries in the charge separation process, which promotes directed energy flow and protects against photodamage \cite{sirohiwal2020protein}. These interactions may help maintain nonpassive states, thus sustaining ergotropic potential under physiological conditions.
Modeling PSIIRC as an open quantum system interacting with structured or unstructured thermal reservoirs that represent the protein environment requires approaches sensitive to coupling strength. These include Markovian and non-Markovian master equations, hierarchical equations of motion, and Green's function-based techniques \cite{Creatore2013,skourtis2010fluctuations,dorfman2013photosynthetic,rouse2024light,PRXEnergy.2.013002,wang2020dissipative,dodin2022noise,poteshman2023network,fang2019nonequilibrium,joubert2023quantum,singh2011electronic,yang2020steady,chen2015using,PhysRevResearch.5.013181,zhang2023many,levi2015quantum,karafyllidis2017quantum,suess2014hierarchy,welack2008single,papp2024computation,sharma2024cotunneling}. These methods clarify the roles of coherence, phonon interaction, and exciton delocalization in shaping the system’s nonequilibrium energy landscape. Experimental setups such as PSIIRC constructs positioned between gold electrodes or metal-coated tips—enable photocurrent measurements and allow direct observation of quantum transport effects \cite{Gerster2012, srinivasan2004experimental}. These findings underscore the potential for probing ergotropy in biological or bio-inspired systems operating out of equilibrium.

In this work, we investigate the ergotropy of PSIIRC under nonequilibrium conditions by analyzing three distinct charge transfer pathways. Each pathway represents a unique configuration of energy flow. By examining the behavior of ergotropy across different excitonic and charge-separated states, we aim to understand how biologically relevant parameters affect the extractable work content of PSIIRC. This approach seeks to bridge quantum thermodynamics with the operational mechanisms of biological energy conversion, offering insights into work extraction in bio-quantum systems. In Sec.~II, we present the PSIIRC Hamiltonian and model the system as a quantum junction using an open-system framework. In Sec.~III, we examine the ergotropy, work, current, and output power for different charge transport pathways and highlight the role of population inversion in ergotropy. Appendices provide quantum chemical details used to realistically simulate the PSIIRC dynamics.

\begin{figure}[ht!]
    \centering
   \includegraphics[width=\columnwidth]{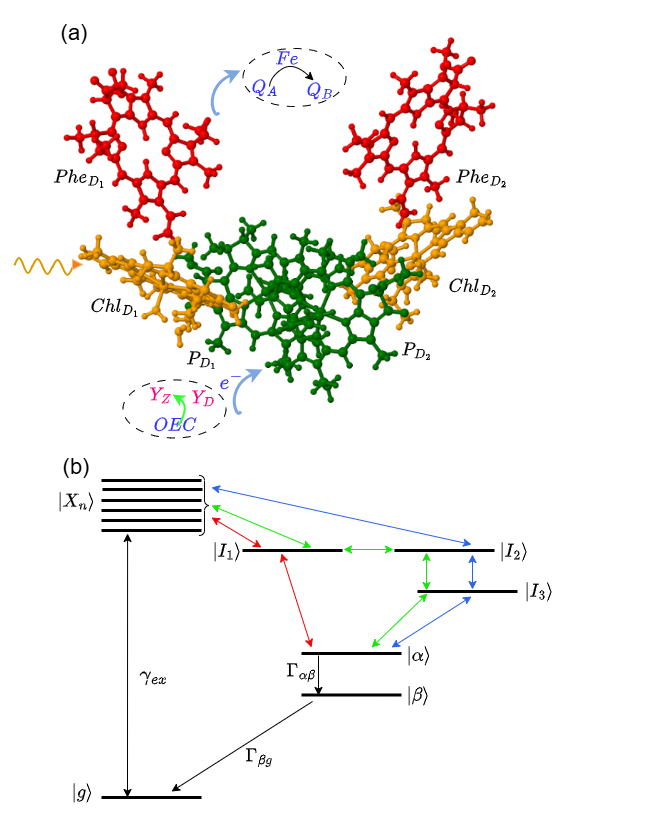}
    \caption{ Model of the PSIIRC as a quantum junction. (a) Schematic depiction of the modeled Photosystem II Reaction Center (PSIIRC). (b) Energy level diagram of the corresponding electronic state space. Single unidirectional arrows denote transition rates from the ground state ($\ket{g}$) and the terminal state ($\ket{\beta}$) to the excitonic manifold. The photoexcitation rate is denoted by $\gamma_{ex}$, while $\Gamma_{\beta g}$ and $\Gamma_{\alpha\beta}$ represent unidirectional transfer rates within the charge transport pathway. [Color online] Red arrows indicate a pathway, which proceeds via the charge-separated state $\ket{I_1}$; blue arrows represent another pathway, involving sequential transitions through $\ket{I_2}$ and $\ket{I_3}$. Green arrows denote a mixed pathway that includes transitions through $\ket{I_1}$, $\ket{I_2}$, and $\ket{I_3} \cite{Novoderezhkin2007}$.
}
    \label{fig:1}  
\end{figure}

\section{\label{sec:level2}PSIIRC Junction}

The Photosystem II Reaction Center (PSIIRC) is a protein-pigment complex that drives the primary photochemical processes in oxygenic photosynthesis. Functionally, it acts as a quantum molecular junction facilitating nonequilibrium electron transport between the Oxygen Evolving Complex (OEC) and the quinone reduction complex (QRC), due to their distinct dynamical timescales \cite{Ferreira2004, Umena2011, Tao2006, STONES20176871}, a schematics of which is shown in Figure \ref{fig:1} (a). Spectroscopic studies reveal that the PSIIRC hosts a ground state $\ket{g}$, six delocalized excitonic states $\ket{X_n}$, four charge-separated states $\ket{I_1}$, $\ket{I_2}$, $\ket{I_3}$, $|\beta\rangle$ and cationic state, $|\alpha\rangle$. Molecular interpretation of these states are known \cite{Brixner2005, Duan2017}. Charge transport proceeds via three distinct pathways involving these states (Figure \ref{fig:1}(b)), shaped by electrostatic shifts and inter-convertive dynamics within the protein environment \cite{Novoderezhkin2011, sirohiwal2020protein}. We present the other relevant details in Appendix A.
Keeping accepted facts in mind, the molecular Hamiltonian of the PSIIRC is written as \cite{STONES20176871,sharma2024cotunneling}: 
\begin{align}
\label{eq-mol-ham}
    \hat{H}_{q} &= E_g \ket{g}\bra{g} 
    + \sum_{n=1}^{6} E_{X_n} \ket{X_n}\bra{X_n}
    + \sum_{m=1}^{3} E_{I_m} \ket{I_m}\bra{I_m} \nonumber \\
    &\quad + E_{\alpha} \ket{\alpha}\bra{\alpha} 
    + E_{\beta} \ket{\beta}\bra{\beta}.
\end{align}
with the overall Hamiltonian being
\begin{equation}\label{A1}
    \hat{H} = \hat{H}_{q} + \hat{H}_{E} + \hat{V}_{\nu}, \quad \hat{V}_{\nu} = \sum_{\nu = b, s, d} \hat{M}_{\nu} \hat{B}_{\nu},
\end{equation}
where $\hat{H}_{q}$ denotes the electronic Hamiltonian of the PSIIRC molecule. The energies of the relevant states are reported in the literature, which we have taken from accepted calculations\cite{STONES20176871}. $\hat{H}_{E}$ corresponds to the environment Hamiltonian, and $\hat{V}_{\nu}$ represents the interaction Hamiltonian between the molecule and the environments. The index $\nu = b, s, d$ refers to the coupling between the molecule and the light bath ($b$), the source ($s$), and the drain ($d$), respectively. Here, $\hat{M}_\nu$ and $\hat{B}_\nu$ are the molecular and environmental operators associated with each terminal.  Note that, the number of states contributing to the dynamics is different in each pathway. Since the system is well known \cite{STONES20176871,sharma2024cotunneling}, we present the complete Hamiltonian, relevant energies and couplings in Appendix A. With this Hamiltonian, a standard quantum master equation can be obtained (Eq. A10 of  appendix A), which is of the form $\ket{\dot{\hat\rho}_q}=\breve {\cal L}\ket{\hat\rho_q}$.  The superoperator represents a rate-matrix with $|\hat\rho_q\rangle =\{\rho_g,\rho_{X_n}^{},\rho_{I_m}^{},\rho_\alpha^{},\rho_\beta^{}\}$ and is constructed by using information from 48 experimentally observed vibrational modes, considering reorganization energies, appropriate line-broadening functions and realistic fluorescence and absorption lineshapes (appendices B and C). The relevant F$\ddot o$rster's rates of charge transfer ($k_{xy}$) are derived by incorporating spectral overlaps between states. The excitonic rates, $r_{xy}$ derived at the Bloch-Redfield levels and the nonequilibrium electron ejection rate $\Gamma_{xy}$ derived perturbatively, between relevant states $x,y\in \hat H_q$ (appendix D). The equation determines the overall time evolution of the reduced density matrix pertaining to the molecular states and is also presented in the appendix A, (Eq.  A10). This master equation serves as the basis for obtained the populations of the states and constructing the active and passive density matrices essential for analysing the ergotropy of the junction.
% and has been well documented \cite{}. 

\section{\label{sec:level4} Ergotropy of the PSIIRC }
The electron transfer between the charged separated state $\ket{\alpha}$ and the cationic state $\ket{\beta}$ serves as an effective terminal where the thermodynamics is of focus \cite{dorfman2013photosynthetic}. The sequential current ($j$), output power ($P$), and useful work ($\mathcal{W}_{\alpha\beta}$) have already been derived at this terminal \cite{STONES20176871}. We focus on the system's ergotropy across three physiologically plausible transport pathways. The first pathway is an exclusive $\ket{I_1}$ pathway that involves photoexcitation to excitonic states $\ket{X_n}$, followed by charge separation to $\ket{I_1}$, and then sequential electron transfer through $\ket{\alpha}$ to $\ket{\beta}$. The second pathway involves two charge transfer intermediate states, which exclusively include the $\ket{I_2}$ and $\ket{I_3}$ states. The  pathway features an alternative route: $\ket{X_n} \rightarrow \ket{I_2} \rightarrow \ket{I_3} \rightarrow \ket{\beta}$. The third pathway is a combined one in which the previous two pathways occur simultaneously, representing a hybrid mechanism that allows population transfer between $\ket{I_1}$, $\ket{I_2}$, and $\ket{I_3}$, before concluding at $\ket{\beta}$. Schematic representation is shown in Figure \ref{fig:1}(b).

To evaluate the ergotropy for each pathway, we begin by expressing the system's Hamiltonian $\hat{H}_q$ in its eigenbasis ordered in a decreasing fashion
\begin{equation}
    E_k \leq E_{k+1}.
\end{equation}
The corresponding steady-state density matrix $\hat{\rho}_q$ is diagonal in this basis. The \textit{passive state} $\hat{\rho}_P$ is obtained by reordering populations in descending order:
\begin{equation}
\label{eq-pop-des}
    P_k \geq P_{k+1},
\end{equation}
where $P_k$ denotes the population of the $k$-th  state and  is obtainable from the master equation, Eq.(A10),
while maintaining the same eigenstates and eigenvalues. The ergotropy $\mathcal{E}$ is then defined as the difference between the expectation values of energy of the active state and that of its passive counterpart:
\begin{align}
    \mathcal{E} &= \text{Tr} \left( \hat{H}_q \hat{\rho}_q \right) - \text{Tr} \left( \hat{H}_q \hat{\rho}_P \right) \\
\label{eq-ergo}
                &= \sum_{k} E_k (\rho_k - P_k).
\end{align}
When the system reaches thermal equilibrium, where $\hat{\rho}_q$ follows the Gibbs distribution, we have $\rho_k = P_k$, and hence $\mathcal{E} = 0$, indicating no extractable work. We evaluate $\mathcal{E}$ as a function of the non-equilibrium electron ejection rate $\Gamma_{\alpha\beta}$. For each pathway, the populations $\rho_k$ are obtained by solving the quantum master equation under non-equilibrium steady-state conditions.  This framework allows us to quantify the pathway-specific maximal extractable work  in the PSIIRC.

\subsection{Exclusive $\ket{I_1}$ pathway}
\begin{figure}[h]
\centering
\includegraphics[width=1\linewidth]{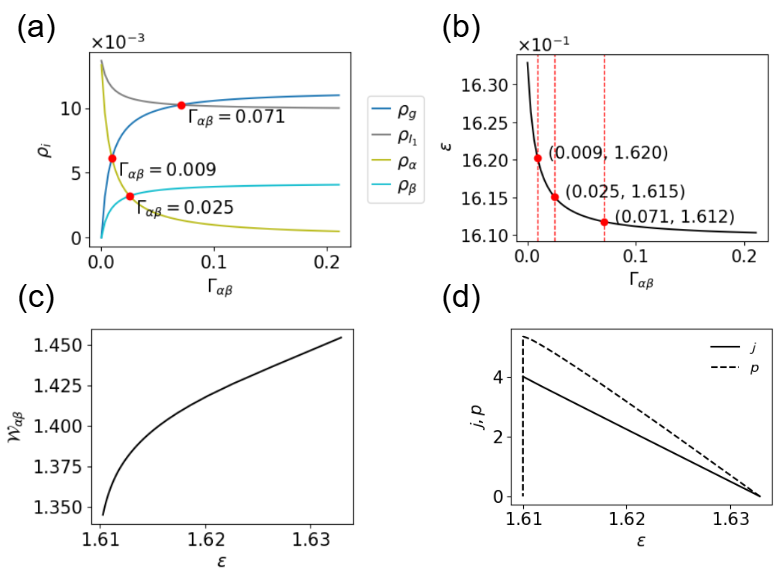}
\caption{Panels (a)–(d) show the variation of system observables with the coupling rate \(\Gamma_{\alpha\beta}\) for exclusive $\ket{I_1}$ pathway. (a) Steady-state populations \(\rho_g\), \(\rho_\beta\), \(\rho_\alpha\), and \(\rho_{I_1}\) as functions of \(\Gamma_{\alpha\beta}\), with population crossovers between respective states clearly indicated. (b) Ergotropy \(\mathcal{E}\) versus \(\Gamma_{\alpha\beta}\) (c),(d) :$\mathcal{W}_{\alpha\beta}{-}\mathcal{E}$, $\mathcal{P}{-}\mathcal{E}$, and $j{-}\mathcal{E}$ for the exclusive $\ket{I_1}$ pathway.
}
\label{fig-path1-erg}
\end{figure}

In the $|I_1\rangle$-exclusive pathway, only the charge-separated state $\ket{I_1}$ is active in the transport cycle. The system Hamiltonian is expressed in ascending energy order as $\{E_g, E_{\beta}, E_{\alpha}, E_{I_1}, E_{X_n}\}$ with $n = 1, \dots, 6$, giving a ten-dimensional Hilbert space. The corresponding steady-state density matrix is diagonal: $\hat{\rho}_i = \text{diag} \{\rho_g, \rho_{\beta}, \rho_{\alpha}, \rho_{I_1}, \rho_{X_n}\}$. These populations are obtained by numerically solving the quantum master equation ($|\dot{\hat{\rho}}\rangle = 0$) and plotting the solutions as a function of the  non-equilibrium electron ejection rate $\Gamma_{\alpha\beta}$ between $\ket{\alpha}$ and $\ket{\beta}$. Figure~\ref{fig-path1-erg}(a) shows the behavior of $\rho_g$, $\rho_{I_1}$, $\rho_{\alpha}$, and $\rho_{\beta}$, while the remaining populations are shown in Figure~\ref{fig-path1-other states} in Appendix A. As $\Gamma_{\alpha\beta}$ increases, three population crossovers occur, yielding four distinct passive states over the full parameter range leading to corresponding changes in the ergotropy. The ergotropy is shown in Figure~\ref{fig-path1-erg}(b). 
Initially, at low $\Gamma_{\alpha\beta}$, the passive state is ordered with $\rho_{X_2}$ and $\rho_{X_5}$ as the highest, followed by $\rho_{I_1}$, $\rho_{\alpha}$, $\rho_g$, and $\rho_{\beta}$, and the rest are ordered by energy. At $\Gamma_{\alpha\beta} = 0.009\,\text{eV}$, $\rho_g$ overtakes $\rho_{\alpha}$, altering the passive state. For $0.025\,\text{eV} < \Gamma_{\alpha\beta} < 0.071\,\text{eV}$, $\rho_{\beta}$ exceeds $\rho_{\alpha}$, prompting another reordering. Finally, beyond $\Gamma_{\alpha\beta} = 0.075\,\text{eV}$, $\rho_g$ exceeds $\rho_{I_1}$, and the passive state stabilizes. Each of these cross-overs mark a change in the dominant terms contributing to the ergotropy. The overall ergotropy curve remains smooth due to the gradual nature of the population shifts, unlike discontinuous behavior seen in quantum engines\cite{PhysRevA.110.032213}.
In the regime $\Gamma_{\alpha\beta} \leq 0.025\,\text{eV}$, the ergotropy is mainly determined by the terms $E_{X_2}(\rho_{X_2}-\rho_{\beta})$, $E_{X_5}(\rho_{X_5}-\rho_{X_3})$, and $E_{\alpha}(\rho_{\alpha}-\rho_{I_1})$. Beyond $\Gamma_{\alpha\beta} = 0.025\,\text{eV}$, $E_{X_2}(\rho_{X_2}-\rho_{\alpha})$ replaces the first term, while $E_{\alpha}(\rho_{\alpha}-\rho_{I_1})$ becomes increasingly dominant. Since $\rho_{\alpha}$ and $\rho_{\beta}$ are of similar magnitude, the ergotropy changes smoothly. These contributions are shown in Figure~\ref{fig-path1-Erg-terms}(a)–(d) of the Appendix section. Other population differences are too small to influence ergotropy significantly. The ergotropy decreases across two regimes due to irreversible energy losses. In the first, for $\Gamma_{\alpha\beta} \leq 0.025\,\text{eV}$, the term $E_{X_2}(\rho_{X_2}-\rho_{\beta})$ decreases by $max(\mathcal{E}) - min(\mathcal{E})=\Delta = -0.00433\,\text{eV}$, while $E_{X_5}(\rho_{X_5}-\rho_{X_3})$ contributes a gain of $\Delta = +0.00076\,\text{eV}$, resulting in a net ergotropy loss of $-0.00357\,\text{eV}$. In the second regime, beyond $\Gamma_{\alpha\beta} = 0.025\,\text{eV}$, the $E_{X_2}(\rho_{X_2}-\rho_{\alpha})$ term contributes a gain of $\Delta = +0.00488\,\text{eV}$, but is outweighed by the fall in $E_{\alpha}(\rho_{\alpha}-\rho_{I_1})$, which decreases by $\Delta = -0.01593\,\text{eV}$, leading to a net drop in extractable work. These shifts reflect how changes in the passive state structure and dominant terms destabilize energy extraction as $\Gamma_{\alpha\beta}$ increases.

Interestingly, when analyzed thermodynamically in terms of ergotropy, this pathway exhibits counter intuitive behavior: as ergotropy (and thus extractable work) increases, the photocurrent decreases, Figure \ref{fig-path1-erg}(c) and (d). While a decreasing behavior of the photocurrent has been previously observed in PSIIRC due to the rate-limiting incoherent transfer from $\ket{X_n}$ to $\ket{I_1}$ under moderate bias \cite{STONES20176871}, here we interpret it in terms of energy-quality trade-offs. Specifically, although the work extracted from the system increases with $\Gamma_{\alpha\beta}$ as seen from Figure \ref{fig-path1-erg}(c), the photocurrent simultaneously drops linearly,  \ref{fig-path1-erg}(d). This suggests a trade-off between transport efficiency and work quality: higher ergotropy coincides with lower $\rho_{\alpha}$ and hence there is a reduced charge mobility. Therefore the PSIIRC behaves like an energy capacitor, storing energy in a more ordered, high-quality form but releasing it less readily. From a thermodynamic perspective, the system shifts toward storing more energy in usable form while dissipating less through charge flow.

% \begin{figure}[h]
% \centering
% \includegraphics[width=1\linewidth]{WJP.png}
% \caption{ (a) and (b):  $\mathcal{W}_{\alpha\beta}{-}\mathcal{E}$, $\mathcal{P}{-}\mathcal{E}$, and $j{-}\mathcal{E}$ for the exclusive $\ket{I_1}$ pathway}

% \label{fig-path1-w,j,p}
% \end{figure}

\subsection{Exclusive $\ket{I_2}$ and $\ket{I_3}$  Pathway}
\begin{figure}[h]
\centering
\includegraphics[width=1\linewidth]{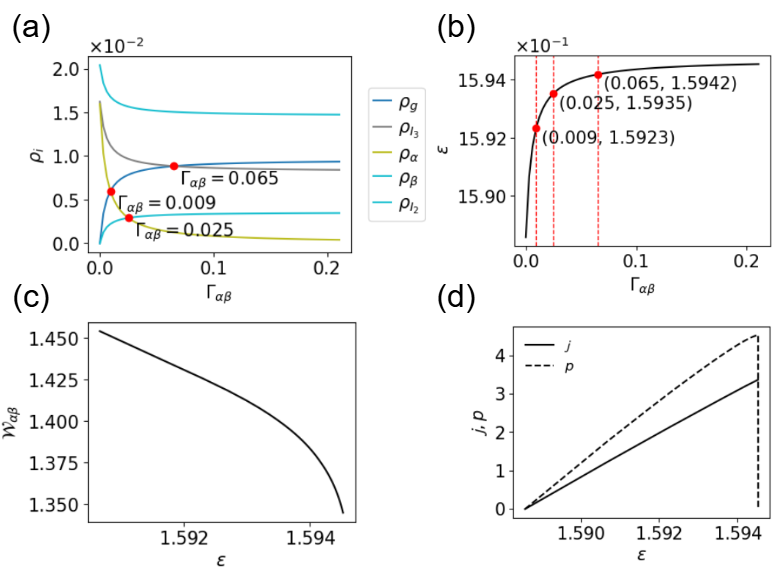}
\caption{Panels (a)–(d) show the variation of system observables with the coupling rate \(\Gamma_{\alpha\beta}\) for exclusive $\ket{I_2}$ and $\ket{I_3}$ pathway. (a) Steady-state populations \(\rho_g\), \(\rho_\beta\), \(\rho_\alpha\), and \(\rho_{I_1}\) as functions of \(\Gamma_{\alpha\beta}\), with population crossovers between respective states clearly indicated. (b) Ergotropy \(\mathcal{E}\) versus \(\Gamma_{\alpha\beta}\) (c),(d) :$\mathcal{W}_{\alpha\beta}{-}\mathcal{E}$, $\mathcal{P}{-}\mathcal{E}$, and $j{-}\mathcal{E}$ for the exclusive $\ket{I_2}$ and $\ket{I_3}$ pathway.
}
\label{fig-path2-erg}

\end{figure}

In the pathway involving charge transfer through $\ket{I_2}$ and $\ket{I_3}$, the system Hamiltonian in the ascending energy basis is given by $H_q = \mathrm{diag}\{E_g, E_\beta, E_\alpha, E_{I_3}, E_{I_2}, E_{X_n}\}$, with $n = 1,\dots,6$. The state $\ket{I_1}$ is excluded, resulting in an eleven-dimensional Hilbert space. The corresponding steady-state density matrix is diagonal: $\hat\rho_i = \mathrm{diag}\{\rho_g, \rho_\beta, \rho_\alpha, \rho_{I_3}, \rho_{I_2}, \rho_{X_n}\}$, with populations $\rho_i$ numerically obtained by solving the quantum master equation ($|\dot{\hat\rho}\rangle = 0$) as a function of the non-equilibrium electron ejection rate $\Gamma_{\alpha\beta}$. As shown in Figure~\ref{fig-path2-erg}(a), three population crossovers occur across $\Gamma_{\alpha\beta}$, resulting in four distinct passive state configurations. The remaining populations are plotted in Figure~\ref{fig-path2-other states} in Appendix A. These define four ergotropy branches, separated by transitions at $\Gamma_{\alpha\beta} \approx 0.009\,\mathrm{eV}$, $0.025\,\mathrm{eV}$, and $0.065\,\mathrm{eV}$ as shown in Figure~\ref{fig-path2-erg}(b). In each interval, the ordering of states in the passive configuration changes. The dominant passive components include $\rho_{X_2}$ and $\rho_{X_5}$, while states like $\rho_g$, $\rho_\alpha$, and $\rho_\beta$ permute depending on $\Gamma_{\alpha\beta}$, leading to corresponding changes in ergotropy.

Within $\Gamma_{\alpha\beta} \leq 0.025\,\mathrm{eV}$, the leading contributors to ergotropy are $E_{X_2}(\rho_{X_2} - \rho_\beta)$, $E_{X_5}(\rho_{X_5} - \rho_{X_3})$, and $E_\alpha(\rho_\alpha - \rho_{I_2})$, resulting in a net gain of $\Delta \mathcal{E} = 0.02076\,\mathrm{eV}$. Beyond this threshold, the first term is replaced by $E_{X_2}(\rho_{X_2} - \rho_\alpha)$, contributing $\Delta = +0.00685\,\mathrm{eV}$, while $E_\alpha(\rho_\alpha - \rho_{I_2})$ drops by $\Delta = -0.01780\,\mathrm{eV}$. The net gain in this regime is $\Delta \approx +0.00981\,\mathrm{eV}$, as shown in Figure~\ref{fig-path2-Erg-terms}(a)–(d). These reflect how reconfigurations in passive states alter energy accessibility.
Figure~\ref{fig-path2-erg}(c) shows that the typical work output initially increases with ergotropy but declines beyond a threshold, while Figure~\ref{fig-path2-erg}(d) reveals that the current increases linearly and the power peaks before falling. Analyzing this behavior with respect to ergotropy reveals a complementary trend compared to the $\ket{I_1}$-exclusive pathway: here, work decreases while current rises. This trade-off highlights a shift in operational mode. In this $\ket{I_2}$-$\ket{I_3}$ pathway, PSIIRC promotes fast charge transport at the expense of energy quality. As $\Gamma_{\alpha\beta}$ increases, $\rho_\alpha$ rises, enhancing carrier mobility but reducing extractable work. Unlike the $\ket{I_1}$ pathway—where higher ergotropy corresponds to better energy storage but slower transport—this regime favors throughput, functioning as a fast-dissipative channel.

This dichotomy illustrates a thermodynamic tension between energy throughput and extractable work. The system stores more non-passive energy, but misalignment between state populations and Hamiltonian energetics limits work conversion. Consequently, the PSIIRC appears to channel energy into fast charge movement, foregoing the structural ordering necessary for optimal work extraction.

% \begin{figure}[h]
% \centering
% \includegraphics[width=1\linewidth]{WJP2.png}
% \caption{ (a) and (b):  $\mathcal{W}_{\alpha\beta}{-}\mathcal{E}$, $\mathcal{P}{-}\mathcal{E}$, and $j{-}\mathcal{E}$ for the exclusive $\ket{I_2}$ and $\ket{I_3}$  Pathway. }

% \label{fig-path2-w,j,p}
% \end{figure}

\subsection{Combined $\ket{I_1}$, $\ket{I_2}$ and $\ket{I_3}$ Pathway}
\begin{figure}[h]
\centering
\includegraphics[width=1\linewidth]{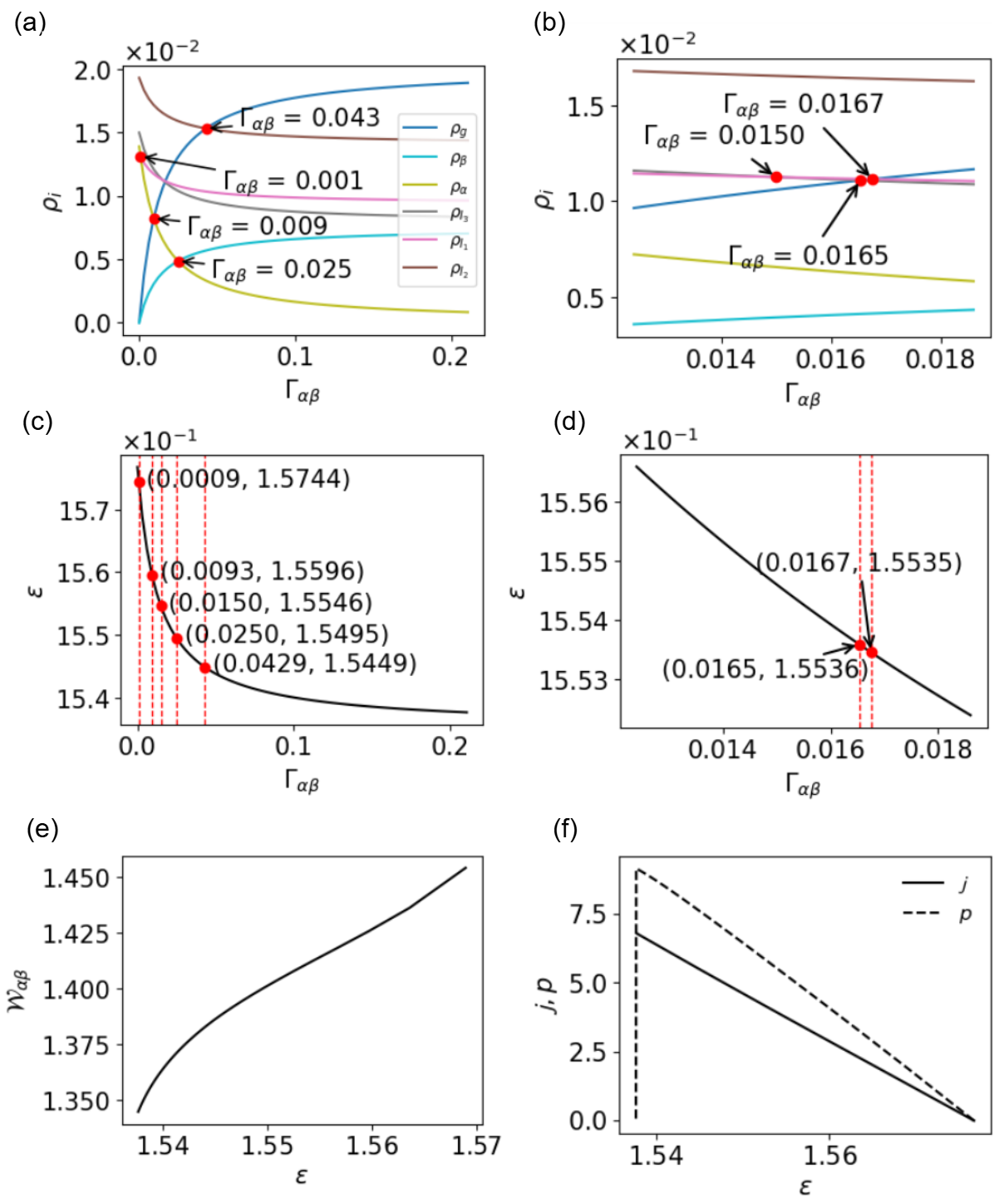}
\caption{Panels (a)–(f) show the variation of system observables with the coupling rate \(\Gamma_{\alpha\beta}\) for the combined $\ket{I_1}$, $\ket{I_2}$ and $\ket{I_3}$ Pathway (a) Steady-state populations \(\rho_g\), \(\rho_\beta\), \(\rho_\alpha\),\(\rho_{I_1}\), \(\rho_{I_2}\) and \(\rho_{I_3}\) as functions of \(\Gamma_{\alpha\beta}\), with population crossovers between respective states clearly indicated. (b) Zoom-in of panel (a), highlighting closely spaced population crossover points as functions of low \(\Gamma_{\alpha\beta}\). (c) Ergotropy \(\mathcal{E}\) versus \(\Gamma_{\alpha\beta}\). (d) Zoom-in of panel (c), resolving nearby ergotropy peak positions and associated \(\Gamma_{\alpha\beta}\) values. (e),(f):$\mathcal{W}_{\alpha\beta}{-}\mathcal{E}$, $\mathcal{P}{-}\mathcal{E}$, and $j{-}\mathcal{E}$ for the combined $\ket{I_1}$, $\ket{I_2}$ and $\ket{I_3}$ pathway.
}
\label{fig-path3-erg}
\end{figure}

In the mixed pathway involving $\ket{I_1}$, $\ket{I_2}$, and $\ket{I_3}$, the system Hamiltonian is defined in the ascending energy basis as $ \mathrm{diag} \left\{ E_g, E_\beta, E_\alpha, E_{I_3}, E_{I_1}, E_{I_2}, E_{X_n} \right\}$, with $n = 1,\dots,6$, resulting in a twelve-dimensional Hilbert space. Multiple population crossovers occur across $\Gamma_{\alpha\beta}$, leading to eight distinct passive configurations and corresponding ergotropy segments, shown in Figure~\ref{fig-path3-erg}(a–b). These transitions occur at seven crossover points up to $\Gamma_{\alpha\beta} = 0.043\,\mathrm{eV}$, beyond which the passive state stabilizes.

The dominant ergotropy contributions before $\Gamma_{\alpha\beta} = 0.025\,\mathrm{eV}$ come from $E_{X_2}(\rho_{X_2} - \rho_\beta)$, $E_{X_5}(\rho_{X_5} - \rho_{X_3})$, and $E_\alpha(\rho_\alpha - \rho_{I_2})$, with a net reduction of $\Delta = -0.00530\,\mathrm{eV}$. Beyond this point, the dominant term shifts to $E_{X_2}(\rho_{X_2} - \rho_\alpha)$, which gains $\Delta = +0.00755\,\mathrm{eV}$, but is offset by a drop of $\Delta = -0.01461\,\mathrm{eV}$ in $E_\alpha(\rho_\alpha - \rho_{I_2})$ as shown in Figure \ref{fig-path3-Erg-terms} in the Appendix, leading to an overall decrease of $\Delta = -0.01236\,\mathrm{eV}$. This profile mirrors the $\ket{I_1}$-exclusive pathway, with similar passive reordering and irreversible energy loss.
Figure~\ref{fig-path3-erg}(c–d) shows that the work increases with ergotropy, while the current decreases linearly. The power exhibits a peak and then falls with increasing ergotropy. These trends indicate that, as in the $\ket{I_1}$-exclusive case, the PSIIRC in this mixed-pathway regime behaves as an energy capacitor, prioritizing the quality of stored energy over rapid transport. Therefore, both the $\ket{I_1}$-exclusive and combined pathways are more suited for extracting high-quality work within the PSIIRC framework.

The major observations of the three case studies highlight the role of the $\ket{I_2}$ charged separated state. Its presence in the second pathway is responsible for the increase in ergotropy but overall causes a decrease in useful extractable work. The absence of contribution from $Chl_{D1}$ and the presence of contribution from $P_{D_2}$(with positive electron density) in the charged separated state of $\ket{I_2}$ is what distinguishes the state from $\ket{I_1}$ and $\ket{I_3}$\cite{STONES20176871}. Therefore the increase in ergotropy and decrease of work in the second pathway can be attributed to the involvement of $P_{D_2}$ in the $\ket{I_2}$ state  allowing the relevant passive states to contribute more to the ergotropy. Note that the magnitude of the ergotropy is the highest for the $\ket{I_1}$ exclusive pathway while the lowest ergotropy is seen for the combined $\ket{I_1}$, $\ket{I_2}$ and $\ket{I_3}$ pathway. This observation may be attributed to a favorable energetic alignment brought about by the contributing chromophores, enhancing  the stability of the charge-
separated intermediates through optimized coupling with
the protein matrix. It is known that the protein environment,
modulates charge-transfer rates and stabilizes specific charge-separated states through electrostatic tuning
and conformational constraints \cite{cupellini2023reaction,lal2021electrostatic}. In particular,
pathways dominated by $Chl_{D1}$ and $Phe_{D1}$ benefit from
enhanced directional energy flow and reduced recombination losses, both necessary for efficient work extraction, leading to large ergotropies. In order to completely quantify the role of the contributing chromophores to the charged separated states in influencing the ergotropy, additional studies and more sophisticated first principle quantum chemical analysis is necessary. We leave it to a future direction of the research.

% These results emphasize that, similar to Pathway 1, \textbf{Pathway 3 is constrained by competing passive states}, and the \textit{abrupt switch} in dominant ergotropy-contributing terms at intermediate $\Gamma_{\alpha\beta}$ values contributes to the \textbf{overall degradation of extractable work}.

% These sequential reconfigurations of the passive state reveal a rich structure of population crossovers that critically influence the ergotropy behavior, especially through the dominant occupations of higher-energy states such as \( \rho_{X_2} \) and \( \rho_{X_5} \). {\color{blue}The ergotropy in the mixed pathway exhibits a decline analogous to that observed in Pathway 1, driven by similar structural transitions and the onset of irreversible dissipative losses.}

% \begin{figure}[h]
% \centering
% \includegraphics[width=1\linewidth]{path3wjp.png}
% \caption{ (a) and (b):  $\mathcal{W}_{\alpha\beta}{-}\mathcal{E}$, $\mathcal{P}{-}\mathcal{E}$, and $j{-}\mathcal{E}$ for the combined I1, I2 and I3 Pathway}

% \label{fig-path3-w,j,p}
% \end{figure}

\section*{Conclusion}

In this study, we conduct a comprehensive theoretical analysis of the Photosystem II  reaction center, employing a master equation approach to model the charge-transfer and excitonic dynamics. The rates of these processes are derived at the Redfield and Förster levels, incorporating realistic spectral profiling to ensure a physically grounded description of the system’s behavior. A central focus of our investigation is the concept of ergotropy, which quantifies the maximum work extractable from a system in a thermodynamically reversible process by computing passive states that represent the system’s non-equilibrium configurations. We further analyze the work load, power, and flux of the PSII reaction center as function of the ergotropy. 

Three primary pathways are considered in our study, each exhibiting distinct ergotropic behaviors. The first pathway involves the charge-separated states of Chl$_{D1}$ and Phe$_{D1}$, while the second pathway includes the charge-separated states of P$_{D1}$ and P$_{D2}$, as well as P$_{D1}$ and Chl$_{D1}$. The third pathway includes all three charge-separated states: Chl$_{D1}$ and Phe$_{D1}$, P$_{D2}$ and P$_{D1}$, and P$_{D1}$ and Chl$_{D1}$. Our findings indicate that the pathway involving only the Chl$_{D1}$ and Phe$_{D1}$ charge-separated states exhibit greater ergotropy, suggesting that the presence of these specific charge-separated states contribute to a more efficient energy conversion process. In contrast, the other pathways, which include additional states, demonstrate lower ergotropy. The presence of Chl$_{D1}$ and Phe$_{D1}$ states probably corresponds to a more favorable energy alignment for charge separation and subsequent energy storage through the passive states. These states may facilitate a more effective redistribution of energy within the system, leading to higher efficiency in work extraction through higher ergotropy. On the other hand, the pathways that involve additional or alternative states are characterized by less favorable energetic conditions, resulting in less efficient energy transfer and lower ergotropy. This behavior is linked to the presence of multi-population crossovers brought about by varying nature of coupling between the contributing chromophores and protein branches. The redistribution of  excitation among the pathways influence the overall energy conversion in a non trivial way through various accessible passive states.

Further, in the pathway involving the charge-separated states Chl$_{D1}$ along with Phe$_{D1}$ and the  pathway  that includes all three charge-separated states: Chl$_{D1}$ and Phe$_{D1}$, P$_{D2}$ and P$_{D1}$, and P$_{D1}$ and Chl$_{D1}$,  favour statistical macrostates capable of acting like energy capacitors, storing energy with higher quality, as evident from the increasing nature of the work and decreasing nature of the photocurrent as a function of ergotropy. In contrast, the pathway involving the charge-separated states of P$_{D1}$ and P$_{D2}$ favour statistical macrostates capable of disbursing energy quickly with lesser quality of work as evident from the decreasing work and increasing photocurrent as the ergotropy keeps increasing. The latter pathway holds charge and energy is present, but not in a very good work-extractable way in comparison to the other two pathways.

These findings shed light on how the reaction center of the photosystem II  leverages its complex network of excitonic states, charge-separated states and cationic states  to optimize energy storage and conversion. By carefully managing the energetic properties of these states, such biosystems can maximize ergotropy and enhance overall energy efficiency. This research deepens our understanding of how quantum thermodynamic principles within the out of equilibrium regime are integrated into the biological processes of photosynthesis.

\appendix

\section{Quantum Master Equation for the PSIIRC}

The Photosystem II Reaction Center (PSII RC) is modeled as a quantum system coupled to multiple environments.  The PSIIRC is a well-known protein-pigment complex responsible for the primary photochemical steps in oxygenic photosynthesis. Its core structure is shown in Figure~(\ref{fig:1}a), based on crystallographic data \cite{Ferreira2004, Umena2011}. Four chlorophyll \textit{a} molecules form a special pair (P$_{D1}$, P$_{D2}$) and an accessory pair (Chl$_{D_1}$, Chl$_{D_2}$). Two pheophytins (Phe$_{D_1}$, Phe$_{D_2}$) and two plastoquinones (Q$_{A}$, Q$_{B}$) are arranged along the D1 and D2 branches \cite{Zouni2001}. Two peripheral chlorophylls, {Chl}$_{Z_{D1}}$ and {Chl}$_{Z_{D2}}$, are located on opposite sides of the reaction center. The PSIIRC is often modeled as a molecular junction situated between the Oxygen Evolving Complex (OEC), serving as an electron source, and the quinone reduction complex (QRC), acting as an electron sink \cite{Tao2006, STONES20176871, sharma2024cotunneling}, since these three structural modules operate on well-separated timescales. This configuration enables directional, nonequilibrium electron transport through the PSIIRC. 
Spectroscopic analyses \cite{Brixner2005, Duan2017} identify the relevant electronic manifold as comprising the ground state $\ket{g}$, six excitonic states $\ket{X_n}$ ($n = 1, \dots, 6$) formed via coherent mixing of the six core chromophores—the four chlorophylls and two pheophytins—and a series of charge-separated states. Among these, $\ket{I_1} \equiv \ket{Chl_{D1}^{+}Phe_{D1}^{-}}$ denotes a charge-separated state between Chl$_{D1}$ and Phe$_{D1}$, $\ket{I_2} \equiv \ket{P_{D2}^{+}P_{D1}^{-}}$ corresponds to charge separation between P$_{D2}$ and P$_{D1}$, and $\ket{I_3} \equiv \ket{P_{D1}^{+}Chl_{D1}^{-}}$ to separation between P$_{D1}$ and Chl$_{D1}$. 
Additional intermediates include $\ket{\alpha} \equiv \ket{P_{D1}^{+}Phe_{D1}^{-}}$, which facilitates stepwise transfer, and $\ket{\beta} \equiv \ket{P_{D1}^{+}Phe_{D1}}$, which serves as a positively charged bridge state influencing downstream charge flow. 
Charge transfer occurs along three principal pathways \cite{Novoderezhkin2011}. In the first, excitonic states $\ket{X_n}$ transition to $\ket{I_1}$ and then to $\ket{\alpha}$. In the second, $\ket{X_n}$ lead to $\ket{I_2}$, then $\ket{I_3}$, and subsequently to $\ket{\beta}$. The third pathway mixes elements of the former two, involving $\ket{I_1}$, $\ket{I_2}$, and $\ket{I_3}$ in a distributed manner. 
Recent theoretical work has shown that the protein matrix induces electrostatic fields that red-shift chlorophyll and blue-shift pheophytin energies, thereby affecting the preferred transfer routes \cite{sirohiwal2020protein}. Singlet-triplet conversions have also been implicated in regulating the charge transfer process \cite{Brixner2005}. Although complete quantum molecular dynamics simulations exist \cite{ogata2013all}, the present study focuses on the dominant charge separation mechanisms relevant for energy conversion within PSIIRC.
Its total Hamiltonian can hence be written as:
\begin{equation}\label{A1}
    \hat{H} = \hat{H}_{\text{q}} + \hat{H}_{\text{E}} + \hat{V}_{\nu}, \quad \hat{V}_{\nu} = \sum_{\nu = b, s, d} \hat{M}_{\nu} \hat{B}_{\nu},
\end{equation}
where $\hat{H}_{\text{q}}$ is the electronic Hamiltonian of the PSII RC, $\hat{H}_{\text{E}}$ represents the environments, and $\hat{V}_{\nu}$ captures the interactions. The index $\nu = b, s, d$ refers to the light bath ($b$), source ($s$), and drain ($d$), respectively, with $\hat{M}_{\nu}$ and $\hat{B}_{\nu}$ denoting the system and environment operators for each channel.

Assuming a separable initial state $\hat{\rho}(0) = \hat{\rho}_{q} \otimes \hat{\rho}_{E}$, and that $\hat{\rho}_{q}$ is a direct product over system components, the reduced density matrix $\hat{\rho}_{q}(t) = \mathrm{Tr}_{E}\{\hat{\rho}(t)\}$ evolves according to:

\begin{align}\label{eq-ide}
    \dot{\widetilde{\rho}}_{q}(t) &= \frac{i^2}{\hbar^{2}} \sum_{i, j = b, s, d} \int_{0}^{t} \left[\widetilde{M}_{i}(t) \widetilde{M}_{j}(t') \widetilde{\rho}_{q}(t) - \widetilde{\rho}_{q}(t) \widetilde{M}_{j}(t') \widetilde{M}_{i}(t)\right] \nonumber \\
    &\quad \times \mathrm{Tr}_{E} \left[\widetilde{B}_{i}(t) \widetilde{B}_{j}(t') \hat{\rho}_{E}(0) - \hat{\rho}_{E}(0) \widetilde{B}_{j}(t') \widetilde{B}_{i}(t)\right] dt',
\end{align}

with interaction picture operators defined as:

\begin{equation}\label{A3}
    \widetilde{O}(t) = e^{\frac{i}{\hbar}(\hat{H}_{q} + \hat{H}_{E})t} \hat{O} e^{-\frac{i}{\hbar}(\hat{H}_{q} + \hat{H}_{E})t}.
\end{equation}
Equation~(\ref{eq-ide}) captures the system-environment correlations: the commutator terms reflect system contributions, while the trace terms define the bath correlation functions~\cite{ritschel2014analytic, timm2008tunneling}. Their structure depends on the system and bath complexity.
For explicit dynamics and correlation evaluations, the molecular Hamiltonian is defined as $\hat{H}_{q} \equiv \hat{H}_{o}$, diagonal in the molecular state basis: ground state $\ket{g}$, excitonic states $\ket{X_n}$, intermediate charge-separated states $\ket{I_{1}}$, $\ket{I_{2}}$, $\ket{I_{3}}$, and long-lived charge-separated states $\ket{\alpha}$ and $\ket{\beta}$.

The environment Hamiltonian consists of three components that describe local environments interacting with the molecule: the radiation field, and the electronic source and drain reservoirs. It is expressed as:

\begin{equation}
    \hat{H}_{E} = \hat{H}_{b} + \hat{H}_{s} + \hat{H}_{d}.
\end{equation}

The radiation field, representing concentrated sunlight, is modeled as a bosonic bath of non-interacting harmonic oscillators. Its Hamiltonian is:

\begin{equation}\label{A6}
    \hat{H}_{b} = \sum_{k} \hbar\omega_{k} \hat{b}_{k}^{\dagger} \hat{b}_{k},
\end{equation}

where $\hat{b}_{k}^{\dagger}$ and $\hat{b}_{k}$ are the bosonic creation and annihilation operators for the $k$th harmonic mode with energy $\hbar \omega_k$.

The source and drain, representing the oxygen-evolving complex (OEC) and plastoquinone reduction complex (QRC), are modeled as fermionic electronic reservoirs. Their Hamiltonian is:

\begin{equation}\label{A7}
    \hat{H}_{e} = \sum_{k \in e} \hbar\widetilde{\omega}_{k e} \hat{c}_{k e}^{\dagger} \hat{c}_{k e}, \quad e = s, d,
\end{equation}
where $\hat{c}_{k e}^{\dagger}$ and $\hat{c}_{k e}$ are fermionic creation and annihilation operators for mode $k$ in the reservoir $e$ (source or drain), with energy $\hbar \widetilde{\omega}_{k e}$. 
For the PSIIRC, the molecule-environment coupling Hamiltonian is composed of three components $(\hat{V} = \hat{V}_{b} + \hat{V}_{s} + \hat{V}_{d})$.  Mathematically,
\begin{align}\label{eq-exc-Ham}
    % \hat{V}_{b} = \hat{M}_{b} \hat{\Gamma}_{b} = \sum_{ki} g_{ki}^{b} \ket{i}\bra{i}(\hat{b}_{k}^{\dagger} + \hat{b}_{k})
    \hat{V}_{b} &= \hat M_{b} \hat B_{b} \equiv \sum_{ki} g_{ki}^{b} \ket{i}\bra{i}(\hat{b}_{k}^{\dagger} + \hat{b}_{k}), ~\ket{i} \in \{\ket {X_{n}} \}.
\end{align}
$g_{ki}^{b}$ represents the effective coupling of the i$^{th}$ molecular electronic state to the k$^{th}$ bosonic mode. The molecule-reservoir coupling Hamiltonian 
     $\hat{V}_{e} = \hat M_{e}\hat B_{e}, ~e =s, d$, based on the spectral evidence and existing reports \cite{Novoderezhkin2011,sharma2024cotunneling,STONES20176871,Creatore2013}, can be explicitly written as,
\begin{align}
\label{eq-fors-ham}
   \hat{V}_{e} &= \sum_{n}^{6} \bigg[t_{I_{1}X_{n}}^{}\hat{c}_{I_{1}}^{\dagger}\hat{c}_{X_n} + \sum_{f=1,3,\beta} t_{\alpha I_{f}}\hat{c}_{\alpha}^{\dagger}\hat{c}_{I_{f}} \nonumber\\
   &+ t_{I_{3}I_{2}}\hat{c}_{I_{3}}^{\dagger}\hat{c}_{I_{2}} + t_{g\beta}^{}\hat{c}_{ks}^{\dagger}\hat{c}_{\beta} + t_{\alpha\beta}^{}\hat{c}_{kd}^{\dagger}\hat{c}_{\alpha} + h.c.\bigg]
\end{align}

with   $t_{mi}$ being the effective coupling term between the i$^{th}$ state to the m$^{th}$ state.

Substituting the explicit forms of the Hamiltonians into Eq.~(\ref{eq-ide}), and applying second- and fourth-order perturbation theory to $\hat{V}_e$ (excluding terms involving $X_n$), we compute the molecular density matrix elements using the projection technique $\rho_{ij} = \bra{i} \hat{\rho}_q \ket{j}$. We define the vectorized density matrix as $\ket{\hat\rho_q} = \{ \rho_{ij} \}$ and derive the master equation in the Schr$\ddot o$dinger picture:

\begin{equation}
    \ket{\dot{\hat\rho}_q} = \breve{\mathcal{L}} \ket{\hat\rho_q},
\end{equation}
where the superoperator $\breve{\mathcal{L}}$ is a $12 \times 12$ Lindblad-type generator with the following explicit structure:
\begin{widetext}
\begin{equation} \label{eq-mas-eq}
\frac{d}{dt}
\left[
    \begin{array}{c}
    \rho_{g} \\[1.5mm]
    \rho_{\beta} \\[1.5mm]
    \rho_{\alpha} \\[1.5mm]
    \rho_{I_{3}} \\[1.5mm]
    \rho_{I_{1}} \\[1.5mm]
    \rho_{I_{2}} \\[1.5mm]
    \rho_{X_{1}} \\[1.5mm]
    \rho_{X_{2}} \\[1.5mm]
    \rho_{X_{3}} \\[1.5mm]
    \rho_{X_{4}} \\[1.5mm]
    \rho_{X_{5}} \\[1.5mm]
    \rho_{X_{6}} 
    \end{array}
\right]
=
\left[
    \begin{array}{cccccccccccc}
    -\gamma_{ex} n & \Gamma_{\beta g} & 0 & 0 & 0 & 0 & \gamma_{ex}(n+1) & 0 & 0 & 0 & 0 & 0  \\[1.5mm]
    0 & \breve{\mathcal{L}}_{\beta\beta} & \Gamma_{\alpha\beta} & 0 & 0 & 0 & 0 & 0 & 0 & 0 & 0 & 0 \\[1.5mm]
    0 & 0 & \breve{\mathcal{L}}_{\alpha\alpha} & k_{I_{3}\alpha} & k_{I_{1}\alpha_{1}} & 0 & 0 & 0 & 0 & 0 & 0 & 0 \\[1.5mm]
    0 & 0 & k_{\alpha I_{3}} & \breve{\mathcal{L}}_{I_3 I_3} & 0 & k_{I_2 I_3} & 0 & 0 & 0 & 0 & 0 & 0 \\[1.5mm]
    0 & 0 & k_{\alpha I_{1}} & 0 & \breve{\mathcal{L}}_{I_1 I_1} & 0 & k_{I_1 1} & k_{I_1 2} & k_{I_1 3} & k_{I_1 4} & k_{I_1 5} & k_{I_1 6} \\[1.5mm]
    0 & 0 & 0 & k_{I_3 I_2} & 0 & \breve{\mathcal{L}}_{I_2 I_2} & k_{I_2 1} & k_{I_2 2} & k_{I_2 3} & k_{I_2 4} & k_{I_2 5} & k_{I_2 6} \\[1.5mm]
    \gamma_{ex} n & 0 & 0 & 0 & k_{1 I_1} & k_{1 I_2} & \breve{\mathcal{L}}_{11} & r_{21} & r_{31} & r_{41} & r_{51} & r_{61} \\[1.5mm]
    0 & 0 & 0 & 0 & k_{2 I_1} & k_{2 I_2} & r_{12} & \breve{\mathcal{L}}_{22} & r_{32} & r_{42} & r_{52} & r_{62} \\[1.5mm]
    0 & 0 & 0 & 0 & k_{3 I_1} & k_{3 I_2} & r_{13} & r_{23} & \breve{\mathcal{L}}_{33} & r_{43} & r_{53} & r_{63} \\[1.5mm]
    0 & 0 & 0 & 0 & k_{4 I_1} & k_{4 I_2} & r_{14} & r_{24} & r_{34} & \breve{\mathcal{L}}_{44} & r_{54} & r_{64} \\[1.5mm]
    0 & 0 & 0 & 0 & k_{5 I_1} & k_{5 I_2} & r_{15} & r_{25} & r_{35} & r_{45} & \breve{\mathcal{L}}_{55} & r_{65} \\[1.5mm]
    0 & 0 & 0 & 0 & k_{6 I_1} & k_{6 I_2} & r_{16} & r_{26} & r_{36} & r_{46} & r_{56} & \breve{\mathcal{L}}_{66}
    \end{array}
\right]
\left[
    \begin{array}{c}
    \rho_{g} \\[1.5mm]
    \rho_{\beta} \\[1.5mm]
    \rho_{\alpha} \\[1.5mm]
    \rho_{I_{3}} \\[1.5mm]
    \rho_{I_{1}} \\[1.5mm]
    \rho_{I_{2}} \\[1.5mm]
    \rho_{X_{1}} \\[1.5mm]
    \rho_{X_{2}} \\[1.5mm]
    \rho_{X_{3}} \\[1.5mm]
    \rho_{X_{4}} \\[1.5mm]
    \rho_{X_{5}} \\[1.5mm]
    \rho_{X_{6}} 
    \end{array}
\right]
\end{equation}
\end{widetext}
For the $\ket{I_1}$ exclusive pathway, the populations of $\ket{I_2}$ and $\ket{I_3}$ are absent, effectively reducing the system's dimensionality to ten. Similarly, for the $I_2, I_3$ exclusive pathway, the population of $\ket{I_1}$ is absent, resulting in an effective dimensional reduction to eleven. To determine the steady states, the quantum master equations with different dimensionalities are solved separately for each case.  Each element of the superoperator $\breve{\mathcal{L}}$ is derived from the evaluation of system-bath correlation functions in Eq.~(\ref{eq-ide}). The matrix satisfies the conditions of detailed balance and probability conservation, which are expressed as:
$
\breve{\mathcal{L}}_{ii} = - \sum_{j \ne i} \breve{\mathcal{L}}_{ji},
$ ensuring that the total population is conserved over time. The rates $\gamma_{ex} n$ and $\gamma_{ex}(n+1)$ arise from the interaction between the molecular system and the electromagnetic field. Here, $\gamma_{ex} \propto |\sum_k g_{kX_1}|^2$ is the excitation rate between the ground state $\ket{g}$ and the excitonic state $\ket{X_1}$, modulated by the thermal occupation $n$ according to the Bose–Einstein distribution. The off-diagonal rates \( r_{pq} \) for \( p \ne q \) describe incoherent transitions within the excitonic manifold and are obtained by evaluating system-bath correlation functions within the Redfield framework. The charge transfer rates \( k_{xy} \), corresponding to transitions between many-body electronic states differing by a single electron, are computed using F\"{o}rster theory~\cite{YANG2002355, runeson2024exciton}.  Finally, the unidirectional nonequilibrium electron transfer rates \( \Gamma_{xy} \) are computed using a standard perturbative approach~\cite{harbola2006quantum}. The methodology for calculating each class of transition rate is detailed in the next section.

The populations $\rho_k$ used in constructing both the original and passive states are obtained by solving the quantum master equation:
\begin{equation}
\label{eq-mas-eq}
    \frac{d}{dt} \hat{\rho}(t) = \mathcal{L} \hat{\rho}(t),
\end{equation}
where $\mathcal{L}$ is the Liouvillian superoperator encoding transitions between electronic states within each pathway.
For each pathway, we numerically obtain steady-state populations $\rho_k$ and rank them to construct the corresponding passive states $\hat{\rho}_P$ via Eq.~(\ref{eq-pop-des}). The quantum chemical data (energy gaps and couplings) for PSIIRC states are used to parameterize the rates in $\mathcal{L}$.

\begin{figure}[h]
\centering
\includegraphics[width=1\linewidth]{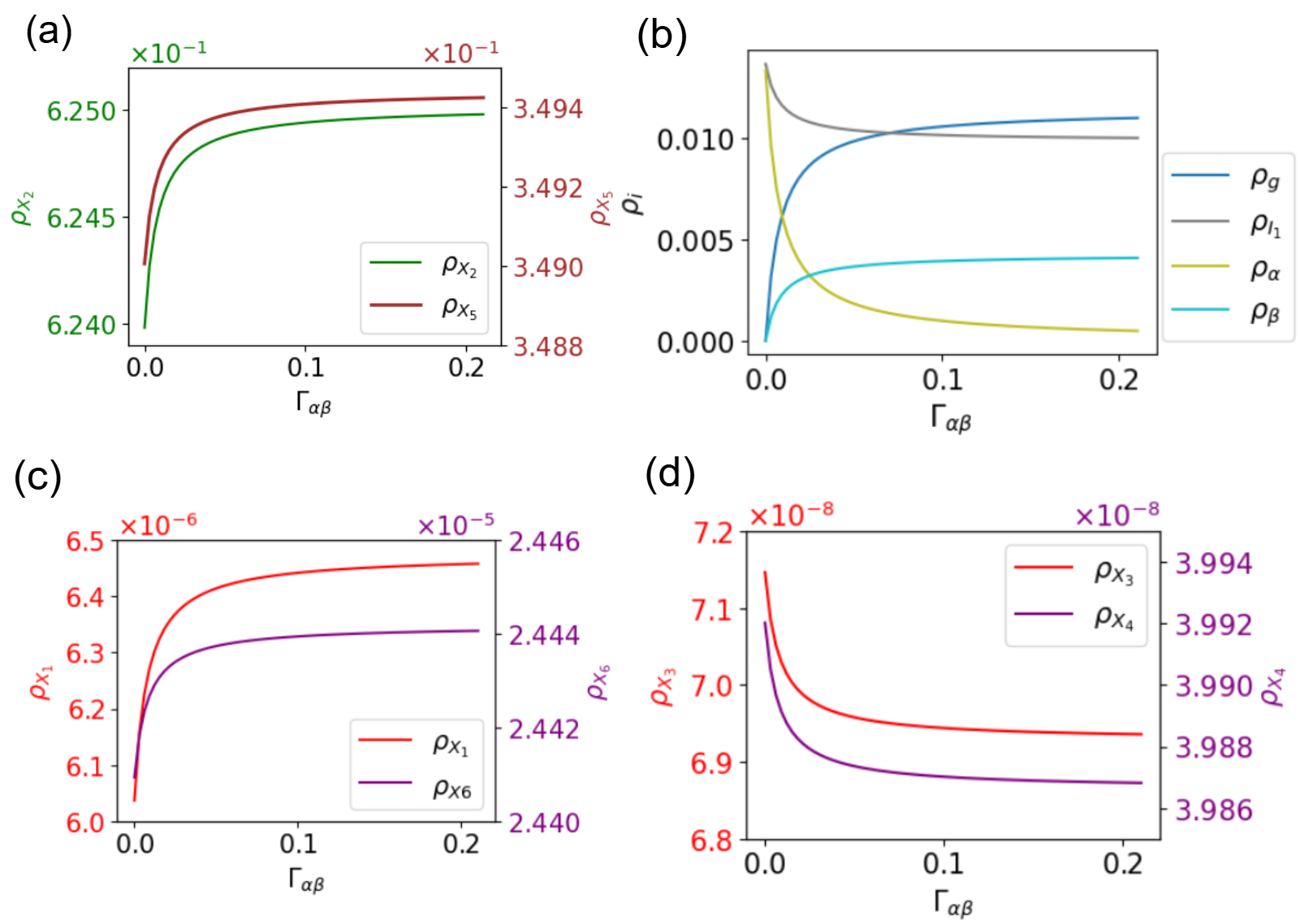}
% \caption{ 
%  Population dynamics of $\rho_{X_1}$,  $\rho_{X_2}$,  $\rho_{X_3}$, $\rho_{X_4}$, $\rho_{X_5}$ and  $\rho_{X_6}$, as functions of $\Gamma_{\alpha\beta}$ for $|I_1\rangle$ exclusive pathway.}
 \caption{ 
 Population dynamics of (a) $\rho_{X_2}$ and $\rho_{X_5}$, (b) $\rho_{g}$, $\rho_{I_1}$, $\rho_{\alpha}$ and $\rho_{\beta}$, (c) $\rho_{X_1}$ and $\rho_{X_6}$, (d) $\rho_{X_3}$ and $\rho_{X_4}$, as functions of $\Gamma_{\alpha\beta}$ for $|I_1\rangle$ exclusive pathway.}

\label{fig-path1-other states}
\end{figure}

\begin{figure}[h]
\centering
\includegraphics[width=1\linewidth]{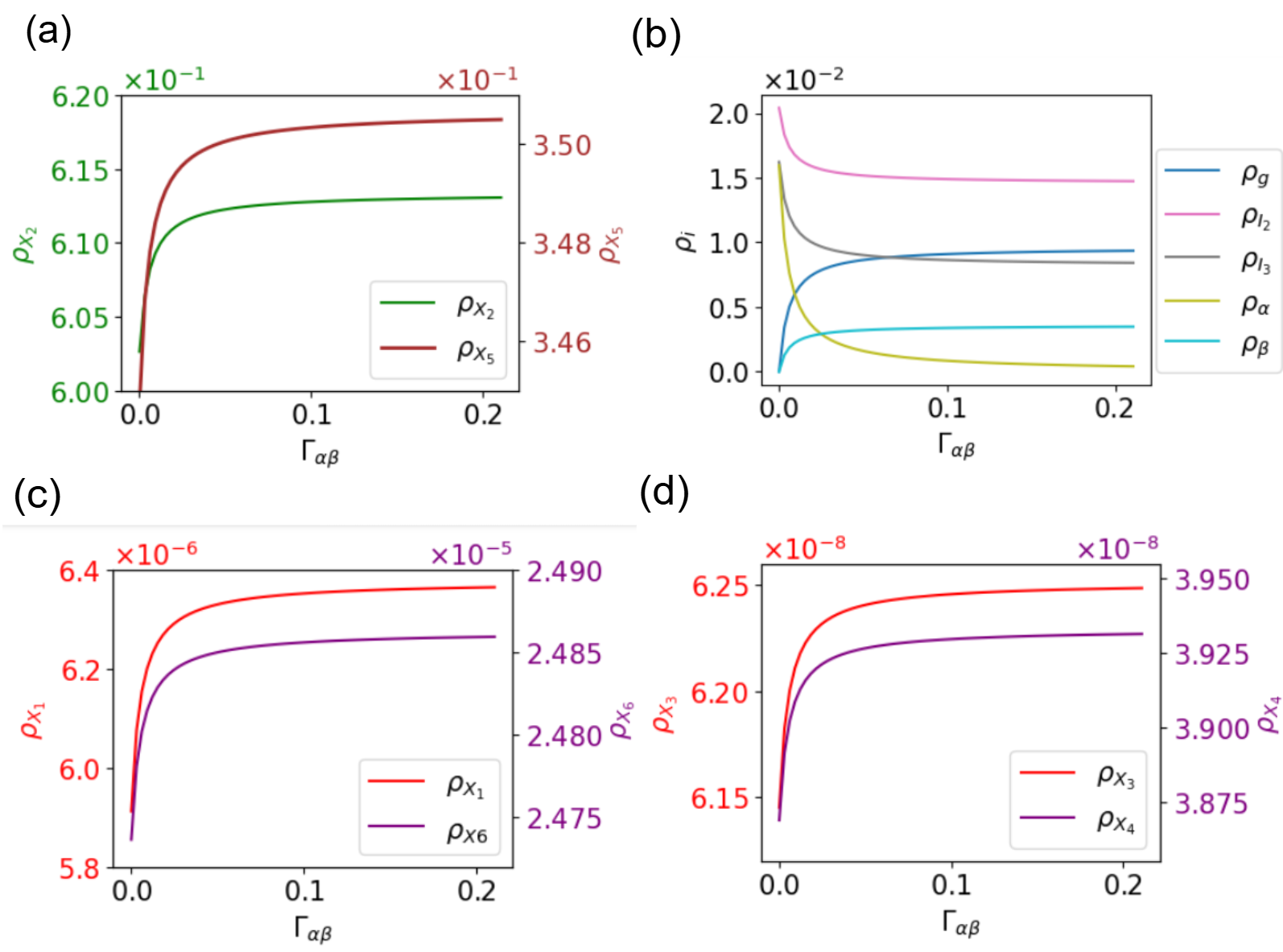}
\caption{ 
 Population dynamics of (a) $\rho_{X_2}$ and $\rho_{X_5}$, (b) $\rho_{g}$, $\rho_{I_2}$, $\rho_{I_3}$, $\rho_{\alpha}$ and $\rho_{\beta}$, (c) $\rho_{X_1}$ and $\rho_{X_6}$, (d) $\rho_{X_3}$ and $\rho_{X_4}$, as functions of $\Gamma_{\alpha\beta}$ for the exclusive $|I_2\rangle,|I_3\rangle$ pathway.}

\label{fig-path2-other states}
\end{figure}

\begin{figure}[h]
\centering
\includegraphics[width=1\linewidth]{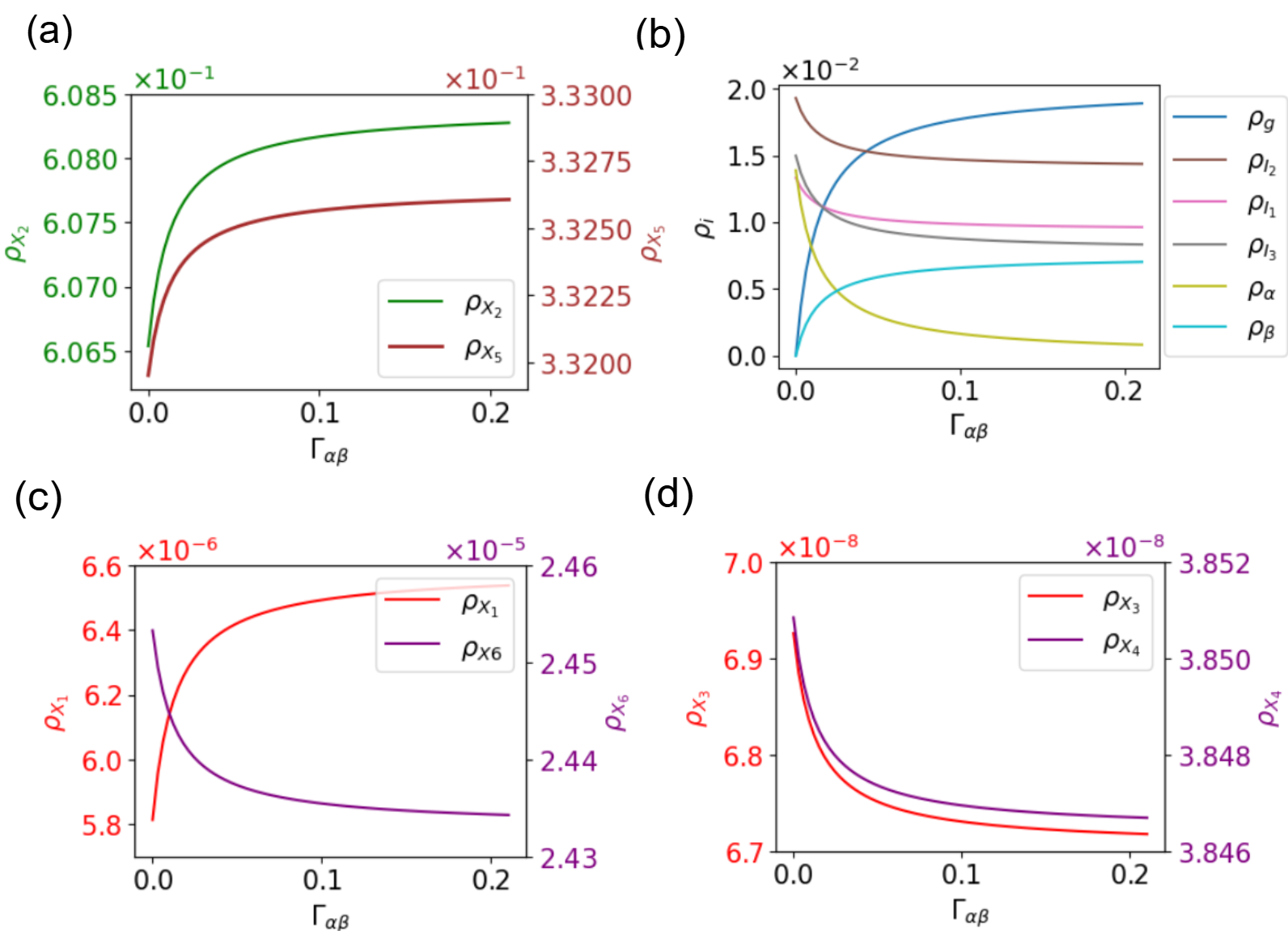}
\caption{
  Population dynamics of (a) $\rho_{X_2}$ and $\rho_{X_5}$, (b) $\rho_{g}$, $\rho_{I_1}$, $\rho_{I_2}$, $\rho_{I_3}$, $\rho_{\alpha}$ and $\rho_{\beta}$, (c) $\rho_{X_1}$ and $\rho_{X_6}$, (d) $\rho_{X_3}$ and $\rho_{X_4}$, as functions of $\Gamma_{\alpha\beta}$ for the combined $\ket{I_1}$, $\ket{I_2}$ and $\ket{I_3}$ pathway.}

\label{fig-path3-other states}
\end{figure}

Across all three pathways, the excitonic states $\rho_{X_2}$ and $\rho_{X_5}$ consistently exhibit the highest populations, as illustrated in Figs.~\ref{fig-path1-other states}, \ref{fig-path2-other states}, and \ref{fig-path3-other states}. The remaining excitonic populations typically follow the hierarchy:
\[
\rho_{X_6} > \rho_{X_1} > \rho_{X_3} > \rho_{X_4},
\]
and this ordering remains largely stable across the full range of $\Gamma_{\alpha\beta}$.
In contrast, the charge-separated states ($\ket{I_1}$, $\ket{I_2}$, $\ket{I_3}$, $\ket{\alpha}$), the terminal state $\ket{\beta}$, and the ground state $\ket{g}$ occupy an intermediate population band. Importantly, the populations among these states permute with varying $\Gamma_{\alpha\beta}$, altering the order required in the passive state construction and thereby changing the ergotropy. These permutation-induced discontinuities in $\hat{\rho}_P$ contribute to the non-monotonic features observed in $\mathcal{E}(\Gamma_{\alpha\beta})$.
Detailed figures (e.g., Figure~\ref{fig-path1-other states},\ref{fig-path2-other states},\ref{fig-path3-other states}) capture these population dynamics and support the claim that pathway-specific mechanisms induce qualitatively different ergotropy profiles.

\begin{figure}[h]
\centering
\includegraphics[width=1\linewidth]{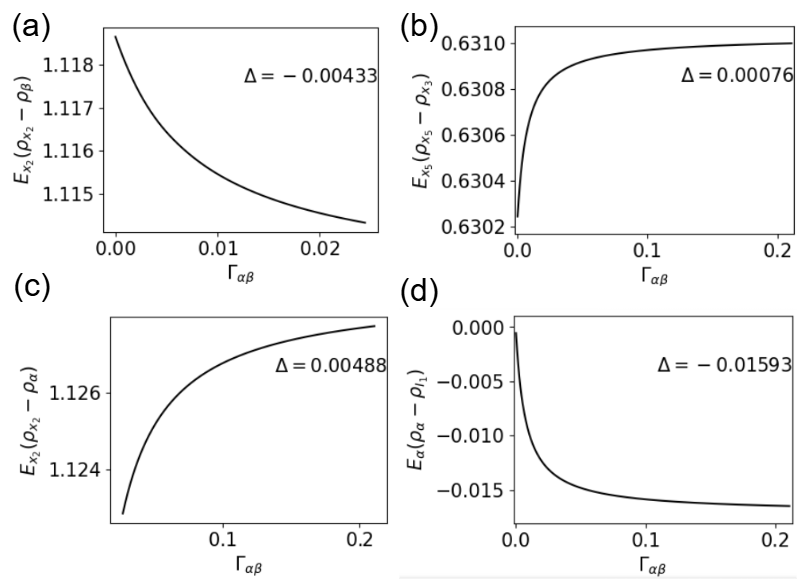}
\caption{(a),(b) :Energy terms \(E_{X_2}(\rho_{X_2} - \rho_\beta)\)  and \(E_{X_5}(\rho_{X_5} - \rho_{X_3})\)  plotted against \(\Gamma_{\alpha\beta}\), with corresponding differences \(\Delta\) shown respectively. (c),(d): Similar plots for \(E_{X_2}(\rho_{X_2} - \rho_\alpha)\) and (f) \(E_\alpha(\rho_\alpha - \rho_{I_1})\) , indicating net variations \(\Delta\) for $|I_1\rangle$ exclusive pathway.}

\label{fig-path1-Erg-terms}
\end{figure}

\begin{figure}[h]
\centering
\includegraphics[width=1\linewidth]{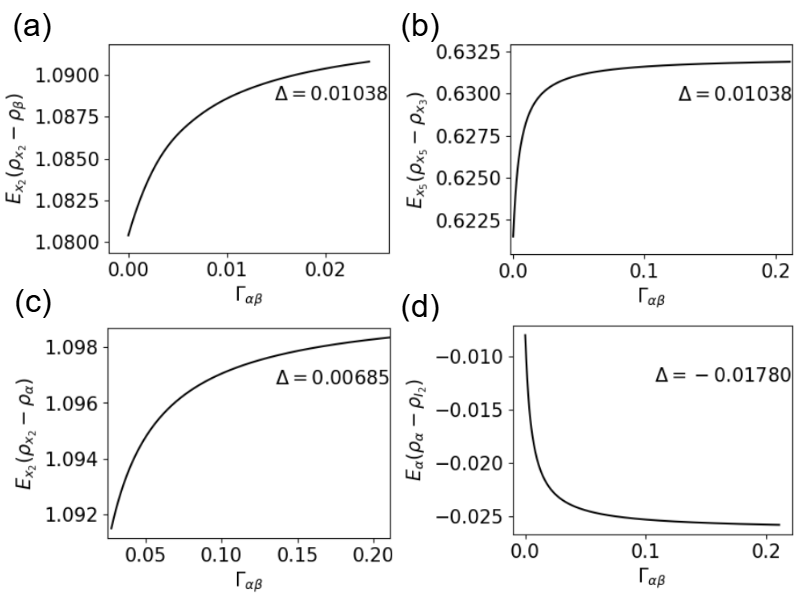}
\caption{(a),(b) :Energy terms \(E_{X_2}(\rho_{X_2} - \rho_\beta)\)  and \(E_{X_5}(\rho_{X_5} - \rho_{X_3})\)  plotted against \(\Gamma_{\alpha\beta}\), with corresponding differences \(\Delta\) shown respectively. (c),(d): Similar plots for \(E_{X_2}(\rho_{X_2} - \rho_\alpha)\) and (f) \(E_\alpha(\rho_\alpha - \rho_{I_1})\) , indicating net variations \(\Delta\) for the exclusive $|I_2\rangle,|I_3\rangle$ pathway.}

\label{fig-path2-Erg-terms}
\end{figure}

\begin{figure}[h]
\centering
\includegraphics[width=1\linewidth]{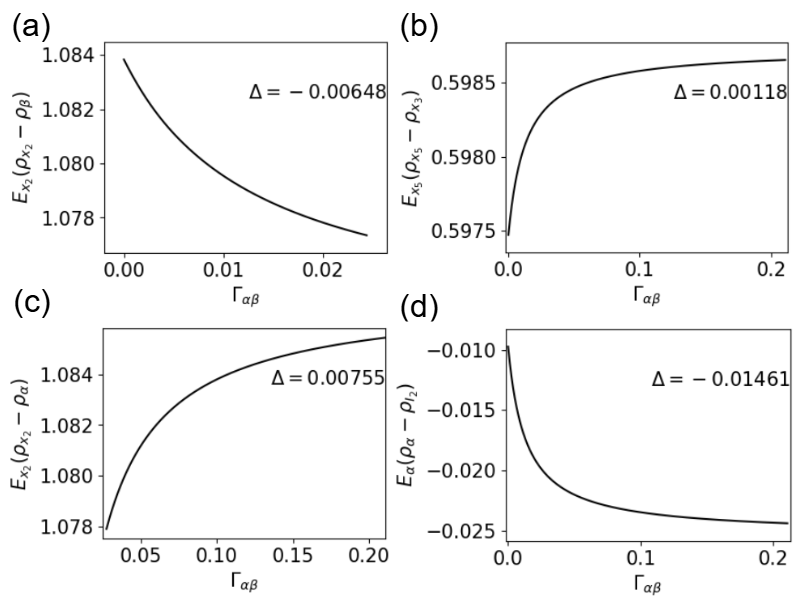}
\caption{(a),(b) :Energy terms \(E_{X_2}(\rho_{X_2} - \rho_\beta)\)  and \(E_{X_5}(\rho_{X_5} - \rho_{X_3})\)  plotted against \(\Gamma_{\alpha\beta}\), with corresponding differences \(\Delta\) shown respectively. (c),(d): Similar plots for \(E_{X_2}(\rho_{X_2} - \rho_\alpha)\) and (f) \(E_\alpha(\rho_\alpha - \rho_{I_1})\) , indicating net variations \(\Delta\) for the combined $\ket{I_1}$, $\ket{I_2}$ and $\ket{I_3}$ Pathway}

\label{fig-path3-Erg-terms}
\end{figure}

\section{Redfield Theory of Excitonic Rates}\label{ED}
Exciton relaxation dynamics govern transitions between quantum states of coupled chromophores, where the total electron number remains conserved across the involved many-body states. These transitions are modeled using a modified Redfield formalism, focusing on population-to-population transfer between excitonic states \( p \) and \( q \), which are elements of the excitonic eigenbasis \( \{ \ket{X_n} \} \). The associated transition rates are derived by inserting the exciton-bath interaction Hamiltonian  into the integro-differential master equation [Eq.~(\ref{eq-ide})], which results in bath correlation functions that mediate the dynamics.

A representative correlation function contributing to the transition between excitonic states \( p \) and \( q \) is given by:
\begin{align}
\langle p| \operatorname{Tr}_E \int_0^t \sum_k g_{kX_n}^b \ket{\tilde{X}_n(t)} \bra{\tilde{X}_n(t')} \tilde{\rho}_q(t) \tilde{b}_k^\dagger(t') \tilde{b}_k(t) \rho_E |q\rangle,
\end{align}
where \( \rho_q(t) = \rho_X(t) \otimes \rho_{q'} \) represents the total density matrix, expressed as the product of the reduced excitonic density matrix \( \rho_X(t) \) and the reduced density matrix of the residual molecular degrees of freedom \( \rho_{q'} \).

Under the assumption of continuous incoherent excitation and dissipative relaxation with well-separated timescales, this expression simplifies to a product form \( r_{pq} \hat{\rho}_{p'q'} \), where \( r_{pq} \) is the excitonic relaxation rate and \( \hat{\rho}_{p'q'} \) is the relevant reduced density matrix element. The explicit form of the transition rate is given by:
\begin{equation} \label{D1}
    r_{pq} = 2 \int_{0}^{\infty} e^{i(\omega_{pq} - \bar{\lambda}) t} e^{-\bar{g}(t)} \eta(t) \, dt,
\end{equation}
where \( \omega_{pq} \) is the energy difference between states \( p \) and \( q \), \( \bar{\lambda} \) is the reorganization energy, \( \bar{g}(t) \) is the lineshape function, and \( \eta(t) \) is the environmental memory kernel governing the bath correlation.

The effective reorganization energy and the difference in line broadening between excitonic states \( p \) and \( q \) are defined as \( \bar{\lambda} = \lambda_{pppp} + \lambda_{qqqq} - 2\lambda_{qqpp} \) and \( \bar{g}(t) = g_{pppp}(t) - g_{qqqq}(t) + 2g_{qqpp}(t) \), respectively. Additionally, a time-dependent term \( \eta(t) = \ddot{g}_{qpqp}(t) - \left( \dot{g}_{qpqq}(t) - \dot{g}_{qppp}(t) - 2i\lambda_{qpqq} \right)^2 \) appears in the transition rate expression. The reorganization energy and line broadening function corresponding to exciton \( p \) are given by \( \lambda_{pppp} = \sum_{i} c_i(p)^4 \lambda_i \) and \( g_{pppp}(t) = \sum_i c_i(p)^4 g_i(t) \), where \( c_i(p) \) is the amplitude of exciton \( p \) on site \( i \), \( \lambda_i \) is the site-specific reorganization energy, and \( g_i(t) \) is the corresponding line broadening function. The excitonic states \( p, q \in \ket{X_n} \), with \( n = 1, 2, \dots, 6 \), are selected such that \( p \ne q \) when calculating the rates. The quantum chemical data required to numerically evaluate these rates are well documented in Olaya-Castro's work \cite{STONES20176871}, which we have reproduced \cite{sharma2024cotunneling} and used in our simulations.

\section{F$\ddot{o}$rster Rates of Charge Transfer }\label{FD}
The transfer rates between charge-separated states, which are many-body states differing by a single electron among the \( N \)-electron configurations, can be estimated using F\"orster theory~\cite{YANG2002355}. This theory yields transfer rates that are equivalent to those predicted by Marcus theory and describes incoherent electron transfer processes. Similar to excitonic transitions, multiple rate terms can be derived by substituting the Hamiltonian in Eq.~(\ref{eq-fors-ham}) into the master equation in Eq.~(\ref{eq-ide}), through the corresponding bath correlation functions. A representative correlation function for transitions between two charge-separated states \( x \) and \( y \) is given as follows:

\begin{align}
\langle x| \operatorname{Tr}_E \int_0^t |t_{I_1X_n}|^2 \tilde{c}_{I_1}^\dag(t) \tilde{c}_{I_1}(t') \tilde{\rho}_q(t) |\tilde{X}_{1}(t') \tilde{X}_{1}(t) \rho_E |y\rangle,
\end{align}
which represents \( k_{xy} \rho_{x'y'} \). The general form of the F\"orster rate expression is approximated as:
\begin{equation}\label{C2}
    k_{xy} = \left| t_{xy} \right|^2 S_{xy}
\end{equation}
where \( t_{xy} \) is the electronic coupling between states \( x \) and \( y \), and \( S_{xy} \) is the spectral overlap between the two states involved in the transfer. The indices \( x \) and \( y \) range over \( 1, 2, 3, 4, 5, 6, I_1, I_2, I_3, \alpha \), and for the F\"orster-like rates and spectral overlap calculations, we assume \( x \neq y \).

The spectral overlap \( S_{xy} \) is expressed as:
\begin{equation} \label{C3}
    S_{xy} = 2 \mathbb{R} \int_{0}^{\infty} dt \, e^{i \omega_{xy} t} e^{-i (\lambda_x + \lambda_y) t - (g_{x}(t) + g_{y}(t))}
\end{equation}
which in the frequency domain becomes:
\begin{equation} \label{C4}
    S_{xy} = \frac{1}{2\pi} \int_{-\infty}^{\infty} d\omega \, \overline{D}_{x}(\omega) D_{y}(\omega)
\end{equation}
where \( \overline{D}_{x}(\omega) = 2 \mathbb{R} \int_{0}^{\infty} dt \, e^{i \omega t} e^{-i \omega_{xy} t + i \lambda_{x} t - g_{x}^{*}(t)} \) and \( \overline{D}_{y}(\omega) = 2 \mathbb{R} \int_{0}^{\infty} dt \, e^{i \omega t} e^{-i \omega_{xy} t - i \lambda_{x} t - g_{x}(t)} \) are the respective fluorescence and absorption lineshapes. The rate \( k_{\beta_2g} \) can be similarly obtained if the spectral profile is known. In the current case, we treat it as a fixed parameter, set to \( 205 \, \text{cm}^{-1} \). The quantum chemical data required to numerically evaluate these rates are well documented in Olaya-Castro's work \cite{STONES20176871}, which we have reproduced \cite{sharma2024cotunneling} and used in our simulations.

% \section{Nonequilibrium Rates}\label{CT}
% The analytical methods of evaluationg the tunneling rates between system and reservoirs  have been studied before using the quantum master framework based on second  perturbation theory. A typical correlation function between two charge-separated states $l$ and $m$ ($l,m\in \ket{\beta}, \ket{g}, \ket{\alpha}$) is,
% \begin{align}
% \langle l|tr_E\displaystyle\int_0^t |t_{g\beta}|^2\tilde c_{g}^\dag(t) \tilde c_{\beta}(t')\tilde\rho_q(t)\tilde c_{d}(t')\tilde c_{d}(t)\rho_E|m\rangle
% \end{align}
% In the above expression, we do a second-order perturbation on the coupling term $\hat V_e$ containing information only on the $l,m$ manybody states. Performing a standard Born-Markov approximation, assuming large bias and completely tracing over the reservoir degree of freedom, the above expression reduces to $\Gamma_{lm}\rho_{l'm'}$, with $\Gamma_{lm}=|t_{\alpha\beta}|^2$. Note that, the Fermi-functions that typically appear in such rates are absent because of the unidirectionality of the carrier electron flow.

\FloatBarrier
\newpage
\bibliographystyle{unsrt}
\bibliography{References}

\begin{thebibliography}{10}

\bibitem{Allahverdyan2004}
Armen~E Allahverdyan, Roger Balian, and Theo~M Nieuwenhuizen.
\newblock Maximal work extraction from finite quantum systems.
\newblock {\em Europhysics Letters}, 67(4):565--571, 2004.

\bibitem{francica2020quantum}
Gianluca Francica, Felix~C Binder, Giacomo Guarnieri, Mark~T Mitchison, John
  Goold, and Francesco Plastina.
\newblock Quantum coherence and ergotropy.
\newblock {\em Physical Review Letters}, 125(18):180603, 2020.

\bibitem{perarnau2015most}
Mart{\'\i} Perarnau-Llobet, Karen~V Hovhannisyan, Marcus Huber, Paul
  Skrzypczyk, Nicolas Brunner, and Antonio Ac{\'\i}n.
\newblock Most energetic passive states.
\newblock {\em Physical Review E}, 92(4):042147, 2015.

\bibitem{vinjanampathy2016quantum}
Sai Vinjanampathy and Janet Anders.
\newblock Quantum thermodynamics.
\newblock {\em Contemporary Physics}, 57(4):545--579, 2016.

\bibitem{alicki2018introduction}
Robert Alicki and Ronnie Kosloff.
\newblock Introduction to quantum thermodynamics.
\newblock {\em arXiv preprint arXiv:1801.08314}, 2018.

\bibitem{binder2015quantum}
Felix~C Binder, Sai Vinjanampathy, Kavan Modi, and John Goold.
\newblock Quantum thermodynamics of local versus global master equations.
\newblock {\em New Journal of Physics}, 17(7):075015, 2015.

\bibitem{andolina2019quantum}
Gian~Marcello Andolina, Maximilian Keck, Andrea Mari, Michele Campisi, Vittorio
  Giovannetti, and Marco Polini.
\newblock Quantum charging advantage cannot be extensive without global
  operations.
\newblock {\em Physical Review Letters}, 122(4):047702, 2019.

\bibitem{joshi2024maximal}
Jitendra Joshi and TS~Mahesh.
\newblock Maximal work extraction unitarily from an unknown quantum state:
  Ergotropy estimation via feedback experiments.
\newblock {\em arXiv preprint arXiv:2409.04087}, 2024.

\bibitem{niu2024experimental}
Zhibo Niu, Yang Wu, Yunhan Wang, Xing Rong, and Jiangfeng Du.
\newblock Experimental investigation of coherent ergotropy in a single spin
  system.
\newblock {\em Physical Review Letters}, 133(18):180401, 2024.

\bibitem{touil2021ergotropy}
Akram Touil, Bar{\i}{\c{s}} {\c{C}}akmak, and Sebastian Deffner.
\newblock Ergotropy from quantum and classical correlations.
\newblock {\em Journal of Physics A: Mathematical and Theoretical},
  55(2):025301, 2021.

\bibitem{PhysRevA.110.032213}
Manash~Jyoti Sarmah and Himangshu~Prabal Goswami.
\newblock Noise-induced coherent ergotropy of a quantum heat engine.
\newblock {\em Phys. Rev. A}, 110:032213, Sep 2024.

\bibitem{allahverdyan2004maximal}
Armen~E Allahverdyan, Roger Balian, and Theo~M Nieuwenhuizen.
\newblock Maximal work extraction from finite quantum systems.
\newblock {\em EPL (Europhysics Letters)}, 67(4):565, 2004.

\bibitem{Ferreira2004}
K.~N. Ferreira, T.~M. Iverson, K.~Maghlaoui, J.~Barber, and S.~Iwata.
\newblock Architecture of the photosystem ii reaction center.
\newblock {\em Science}, 303:1831, 2004.

\bibitem{blankenship2014molecular}
Robert~E Blankenship.
\newblock {\em Molecular mechanisms of photosynthesis}.
\newblock John Wiley \& Sons, 2014.

\bibitem{scholes2011lessons}
Gregory~D Scholes, Graham~R Fleming, Alexandra Olaya-Castro, and Rienk van
  Grondelle.
\newblock Lessons from nature about solar light harvesting.
\newblock {\em Nature Chemistry}, 3(10):763--774, 2011.

\bibitem{van2000photosynthetic}
Herbert van Amerongen, Leonas Valkunas, and Rienk van Grondelle.
\newblock Photosynthetic excitation transfer: Theoretical and experimental
  insights.
\newblock {\em Science}, 289(5485):943--948, 2000.

\bibitem{engel2007evidence}
Gregory~S Engel, Tessa~R Calhoun, Elizabeth~L Read, Tae-Kyu Ahn,
  Tom{\'a}{\v{s}} Man{\v{c}}al, Yuan-Chung Cheng, Robert~E Blankenship, and
  Graham~R Fleming.
\newblock Evidence for wavelike energy transfer through quantum coherence in
  photosynthetic systems.
\newblock {\em Nature}, 446(7137):782--786, 2007.

\bibitem{STONES20176871}
Richard Stones, Hoda Hossein-Nejad, Rienk {van Grondelle}, and Alexandra
  Olaya-Castro.
\newblock On the performance of a photosystem ii reaction centre-based
  photocell††electronic supplementary information (esi) available. see doi:
  10.1039/c7sc02983g.
\newblock {\em Chemical Science}, 8(10):6871--6880, 2017.

\bibitem{Novoderezhkin2011}
V.~I. Novoderezhkin, E.~Romero, J.~P. Dekker, and R.~van Grondelle.
\newblock Multiple charge-separation pathways in photosystem ii: Modeling of
  transient absorption kinetics.
\newblock {\em ChemPhysChem}, 12(4):681--688, 2011.

\bibitem{ishizaki2009theoretical}
Akihito Ishizaki and Graham~R Fleming.
\newblock Theoretical examination of quantum coherence in a photosynthetic
  system at physiological temperature.
\newblock {\em Proceedings of the National Academy of Sciences},
  106(41):17255--17260, 2009.

\bibitem{hoki2009quantum}
Kenji Hoki and Paul Brumer.
\newblock Quantum-mechanical nature of electronic transitions in photosynthetic
  light harvesting.
\newblock {\em The Journal of Physical Chemistry B}, 113(45):15791--15800,
  2009.

\bibitem{fassioli2010quantum}
Francesca Fassioli, Alexandra Olaya-Castro, Simon Scheuring, James~N Sturgis,
  and Rienk van Grondelle.
\newblock Quantum coherence, energy transfer and photosynthesis.
\newblock {\em New Journal of Physics}, 12(8):085005, 2010.

\bibitem{sirohiwal2020protein}
Abhishek Sirohiwal, Frank Neese, and Dimitrios~A Pantazis.
\newblock Protein matrix control of reaction center excitation in photosystem
  {II}.
\newblock {\em Journal of the American Chemical Society}, 142(42):18174--18190,
  2020.

\bibitem{Creatore2013}
C.~Creatore, M.~Parker, S.~Emmott, and A.~Chin.
\newblock Efficient biologically inspired photocell enhanced by quantum
  coherence.
\newblock {\em Physical Review Letters}, 111:253601, 2013.

\bibitem{skourtis2010fluctuations}
Spiros~S Skourtis, David~H Waldeck, and David~N Beratan.
\newblock Fluctuations in biological and bioinspired electron-transfer
  reactions.
\newblock {\em Annual Reviews of Physical Chemistry}, 61(1):461--485, 2010.

\bibitem{dorfman2013photosynthetic}
Konstantin~E Dorfman, Dmitri~V Voronine, Shaul Mukamel, and Marlan~O Scully.
\newblock Photosynthetic reaction center as a quantum heat engine.
\newblock {\em Proceedings of the National Academy of Sciences},
  110(8):2746--2751, 2013.

\bibitem{rouse2024light}
Dominic~M Rouse, Adesh Kushwaha, Stefano Tomasi, Brendon~W Lovett, Erik~M
  Gauger, and Ivan Kassal.
\newblock Light-harvesting efficiency cannot depend on optical coherence in the
  absence of orientational order.
\newblock {\em The Journal of Physical Chemistry Letters}, 15(1):254--261,
  2024.

\bibitem{PRXEnergy.2.013002}
Nicholas Werren, Will Brown, and Erik~M. Gauger.
\newblock Light harvesting enhanced by quantum ratchet states.
\newblock {\em PRX Energy}, 2:013002, Feb 2023.

\bibitem{wang2020dissipative}
Zibo Wang and Imran Mirza.
\newblock Dissipative five-level quantum systems: A quantum model of
  photosynthetic reaction centers.
\newblock pages JM6B--26. Optica Publishing Group, 2020.

\bibitem{dodin2022noise}
Amro Dodin and Paul Brumer.
\newblock Noise-induced coherence in molecular processes.
\newblock {\em Journal of Physics B: Atomic, Molecular and Optical Physics},
  54(22):223001, 2022.

\bibitem{poteshman2023network}
Abigail~N Poteshman, Mathieu Ouellet, Lee~C Bassett, and Dani~S Bassett.
\newblock Network structure and dynamics of effective models of nonequilibrium
  quantum transport.
\newblock {\em Physical Review Research}, 5(2):023125, 2023.

\bibitem{fang2019nonequilibrium}
Xiaona Fang, Karsten Kruse, Ting Lu, and Jin Wang.
\newblock Nonequilibrium physics in biology.
\newblock {\em Reviews of Modern Physics}, 91(4):045004, 2019.

\bibitem{joubert2023quantum}
Lo{\"\i}c Joubert-Doriol, Kenneth~A Jung, Artur~F Izmaylov, and Paul Brumer.
\newblock Quantum kinetic rates within the nonequilibrium steady state.
\newblock {\em Journal of Chemical Theory and Computation}, 19(4):1130--1143,
  2023.

\bibitem{singh2011electronic}
Navinder Singh and Paul Brumer.
\newblock Electronic energy transfer in model photosynthetic systems: Markovian
  vs. non-markovian dynamics.
\newblock {\em Faraday Discussions}, 153:41--50, 2011.

\bibitem{yang2020steady}
Pei-Yun Yang and Jianshu Cao.
\newblock Steady-state analysis of light-harvesting energy transfer driven by
  incoherent light: From dimers to networks.
\newblock {\em The Journal of Physical Chemistry Letters}, 11(17):7204--7211,
  2020.

\bibitem{chen2015using}
Hong-Bin Chen, Neill Lambert, Yuan-Chung Cheng, Yueh-Nan Chen, and Franco Nori.
\newblock Using non-markovian measures to evaluate quantum master equations for
  photosynthesis.
\newblock {\em Scientific reports}, 5(1):12753, 2015.

\bibitem{PhysRevResearch.5.013181}
Neill Lambert, Tarun Raheja, Simon Cross, Paul Menczel, Shahnawaz Ahmed,
  Alexander Pitchford, Daniel Burgarth, and Franco Nori.
\newblock Qutip-bofin: A bosonic and fermionic numerical
  hierarchical-equations-of-motion library with applications in
  light-harvesting, quantum control, and single-molecule electronics.
\newblock {\em Phys. Rev. Res.}, 5:013181, Mar 2023.

\bibitem{zhang2023many}
Min Zhang, Yaru Liu, Ya-nan Jiang, and Yuchen Ma.
\newblock Many-body green’s function theory for electronic excitations in
  complex chemical systems.
\newblock {\em The Journal of Physical Chemistry Letters}, 14(23):5267--5282,
  2023.

\bibitem{levi2015quantum}
Federico Levi, Stefano Mostarda, Francesco Rao, and Florian Mintert.
\newblock Quantum mechanics of excitation transport in photosynthetic
  complexes: a key issues review.
\newblock {\em Reports on Progress in Physics}, 78(8):082001, 2015.

\bibitem{karafyllidis2017quantum}
Ioannis~G Karafyllidis.
\newblock Quantum transport in the fmo photosynthetic light-harvesting complex.
\newblock {\em Journal of Biological Physics}, 43:239--245, 2017.

\bibitem{suess2014hierarchy}
D~Suess, A~Eisfeld, and WT~Strunz.
\newblock Hierarchy of stochastic pure states for open quantum system dynamics.
\newblock {\em Physical Review Letters}, 113(15):150403, 2014.

\bibitem{welack2008single}
Sven Welack, Jeremy~B Maddox, Massimiliano Esposito, Upendra Harbola, and Shaul
  Mukamel.
\newblock Single-electron counting spectroscopy: simulation study of porphyrin
  in a molecular junction.
\newblock {\em Nano letters}, 8(4):1137--1141, 2008.

\bibitem{papp2024computation}
Eszter Papp and G{\'a}bor Vattay.
\newblock Computation of biological conductance with liouville quantum master
  equation.
\newblock {\em Scientific Reports}, 14(1):19571, 2024.

\bibitem{sharma2024cotunneling}
Debasish Sharma, Manash~Jyoti Sarmah, Mriganka Sandilya, and Himangshu~Prabal
  Goswami.
\newblock Cotunneling assisted nonequilibrium thermodynamics of a
  photosynthetic junction.
\newblock {\em The Journal of Chemical Physics}, 162(24):245101, 06 2025.

\bibitem{Gerster2012}
Daniel Gerster, Joachim Reichert, Hai Bi, Johannes~V. Barth, Simone~M. Kaniber,
  Alexander~W. Holleitner, Iris Visoly-Fisher, Shlomi Sergani, and Itai
  Carmeli.
\newblock Photocurrent of a single photosynthetic protein.
\newblock {\em Nature Nanotechnology}, 7(10):673--676, 2012.

\bibitem{srinivasan2004experimental}
Narayan Srinivasan, Marcel Wendling, and Rienk van Grondelle.
\newblock Experimental studies of psiirc photocurrents.
\newblock {\em Journal of Photochemistry and Photobiology B: Biology},
  76(1-3):179--189, 2004.

\bibitem{Novoderezhkin2007}
V.~I. Novoderezhkin, J.~P. Dekker, and R.~Van Grondelle.
\newblock Dynamics of excitation energy transfer in the photosystem ii core
  complex.
\newblock {\em Biophysical Journal}, 93:1293, 2007.

\bibitem{Umena2011}
Y.~Umena, K.~Kawakami, J.-R. Shen, and N.~Kamiya.
\newblock Crystal structure of oxygen-evolving photosystem ii at a resolution
  of 1.9 Å.
\newblock {\em Nature}, 473:55, 2011.

\bibitem{Tao2006}
N.~J. Tao.
\newblock Electron transport in molecular junctions.
\newblock {\em Nature Nanotechnology}, 1:173, 2006.

\bibitem{Brixner2005}
T.~Brixner, J.~Stenger, H.~M. Vaswani, M.~Cho, R.~E. Blankenship, and G.~R.
  Fleming.
\newblock Two-dimensional spectroscopy of electronic couplings in
  photosynthesis.
\newblock {\em Nature}, 434:625, 2005.

\bibitem{Duan2017}
H.-G. Duan, V.~I. Prokhorenko, E.~Wientjes, R.~Croce, M.~Thorwart, and R.~J.~D.
  Miller.
\newblock Nature’s light-harvesting engine: A quantum-coherent energy
  transport network.
\newblock {\em Scientific Reports}, 7:12347, 2017.

\bibitem{cupellini2023reaction}
Lorenzo Cupellini, Christian Teutloff, Jan Pieper, Frank Müh, Thomas Renger,
  Stefano Jurinovich, and Benedetta Mennucci.
\newblock Reaction center excitation in photosystem ii: From multiscale
  modeling to functional principles.
\newblock {\em Accounts of Chemical Research}, 56(21):1592--1603, 2023.

\bibitem{lal2021electrostatic}
Shubhadeep Lal, Pavel Pospíšil, and M.~Durga Prasad.
\newblock Electrostatic profiling of photosynthetic pigments: implications for
  directed spectral tuning.
\newblock {\em Physical Chemistry Chemical Physics}, 23(35):19830--19841, 2021.

\bibitem{Zouni2001}
A.~Zouni, H.-T. Witt, J.~Kern, P.~Fromme, N.~Krauss, W.~Saenger, and P.~Orth.
\newblock Crystal structure of photosystem ii from synechococcus elongatus at
  3.8 Å resolution.
\newblock {\em Nature}, 409:739, 2001.

\bibitem{ogata2013all}
Koji Ogata, Taichi Yuki, Makoto Hatakeyama, Waka Uchida, and Shinichiro
  Nakamura.
\newblock All-atom molecular dynamics simulation of photosystem ii embedded in
  thylakoid membrane.
\newblock {\em Journal of the American Chemical Society}, 135(42):15670--15673,
  2013.

\bibitem{ritschel2014analytic}
Gerhard Ritschel and Alexander Eisfeld.
\newblock An analytic continuation approach to nonequilibrium quantum
  transport.
\newblock {\em Journal of Chemical Physics}, 141(9):094101, 2014.

\bibitem{timm2008tunneling}
Carsten Timm.
\newblock Tunneling through molecules and quantum dots: Master-equation
  approaches.
\newblock {\em Physical Review B}, 77(19):195416, 2008.

\bibitem{YANG2002355}
Mingli Yang and Graham~R. Fleming.
\newblock A quantum theory of exciton--electron transfer kinetics.
\newblock {\em Chemical Physics}, 282(2--3):355--368, 2002.

\bibitem{runeson2024exciton}
Jesper~E. Runeson et~al.
\newblock Exciton transfer in donor-bridge-acceptor molecular systems: A
  unified treatment of coherent and incoherent dynamics.
\newblock {\em The Journal of Chemical Physics}, 160(5):054105, 2024.

\bibitem{harbola2006quantum}
Upendra Harbola, Massimiliano Esposito, and Shaul Mukamel.
\newblock Quantum master equation for electron transport through quantum dots
  and single molecules.
\newblock {\em Physical Review B}, 74(23):235309, 2006.

\end{thebibliography}

\end{document}